\DocumentMetadata{}
\documentclass[sigplan,10pt,nonacm]{acmart}
\usepackage{xurl}
\usepackage[utf8]{inputenc}
\usepackage{wasysym}
\usepackage{xspace}
\usepackage{graphicx}
\usepackage{subcaption}
\usepackage{xcolor}

\newcommand{\system}{GlucOS\xspace}
\newcommand{\kernel}{BioKernel\xspace}

\newcommand{\caleb}[1]{\textcolor{red}{\textbf{[ Caleb: #1]}}}

\newcommand{\validate}{validate\xspace}
\begin{document}

\title{\system: Security, correctness, and simplicity for automated insulin delivery}
\author{Hari Venugopalan}
\affiliation{
\institution{University of California, Davis}
\city{Davis}
\country{USA}
}
\email{hvenugopalan@ucdavis.edu}

\author{Shreyas Madhav Ambattur Vijayanand}
\affiliation{
\institution{University of California, Davis}
\city{Davis}
\country{USA}
}
\email{smvijay@ucdavis.edu}

\author{Caleb Stanford}
\affiliation{
\institution{University of California, Davis}
\city{Davis}
\country{USA}
}
\email{cdstanford@ucdavis.edu}

\author{Stephanie Crossen}
\affiliation{
\institution{University of California, Davis}
\city{Davis}
\country{USA}
}
\email{scrossen@ucdavis.edu}

\author{Samuel T. King}
\affiliation{
\institution{University of California, Davis}
\city{Davis}
\country{USA}
}
\email{kingst@ucdavis.edu}

\begin{abstract}
We present \system, a novel system for trustworthy automated insulin
delivery. Fundamentally, this paper is about a system we designed,
implemented, and deployed on real humans and the lessons learned from
our experiences.

\system introduces a novel architecture that allows users to personalize
diabetes management using any predictive model (including ML) for insulin
dosing while simultaneously protecting them against malicious models. We
also introduce a novel holistic security mechanism that adapts to 
unprecedented changes to human physiology. We use formal methods to prove
correctness of critical components and incorporate humans as part of our
defensive strategy. Our evaluation includes both a real-world deployment
with seven individuals and results from simulation to show that our
techniques generalize. We highlight that our results are \emph{not} from a
lab study, with people using \system to manage Type 1 Diabetes in their
daily lives. Our results show that \system maintains safety and improves
glucose control even under attack conditions. This work demonstrates the
potential for secure, personalized, automated healthcare systems. Our entire
source code is available at \href{https://anonymous.4open.science/r/metabolicos-02B7/} {this link}.
\end{abstract}

\maketitle
\pagestyle{plain}

\section{Introduction}
\label{sec:intro}
\begin{figure*}[t]
  \centering
  \begin{subfigure}[b]{0.2\textwidth}
    \centering
    \includegraphics[height=2.5cm]{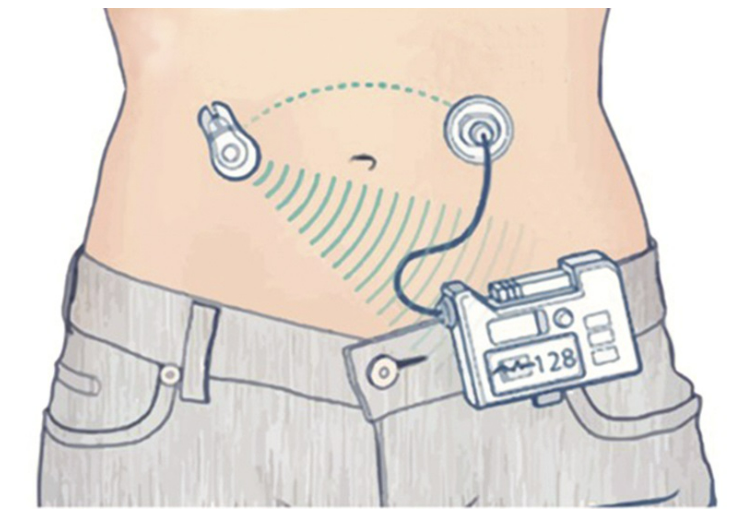}
    \caption{Devices for automated insulin delivery. This figure is from breakthrought1d.org.}
    \label{fig:devices}
  \end{subfigure}
  \quad
  \begin{subfigure}[b]{0.32\textwidth}
    \centering
    \includegraphics[height=4cm]{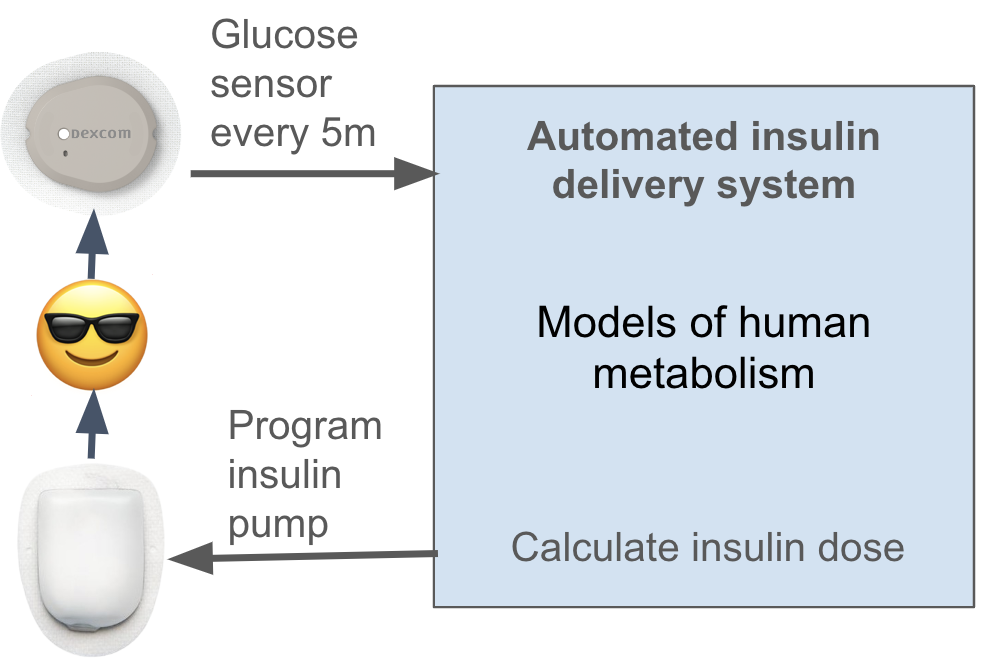}
    \caption{Closed-loop architecture.}
    \label{fig:closed_loop}
  \end{subfigure}
  \quad
  \begin{subfigure}[b]{0.38\textwidth}
    \centering
    \includegraphics[height=4cm]{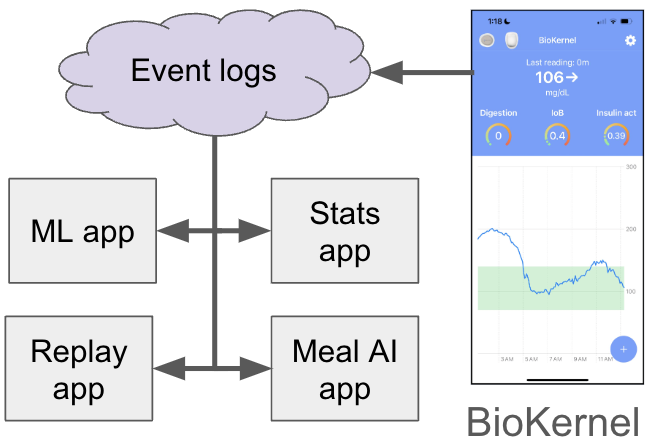}
    \caption{Overall \system architecture.}
    \label{fig:biokernel2}
  \end{subfigure}
  \caption{Overview of automated insulin delivery devices, closed-loop
    architecture, and \system.}
  \hrulefill
  \label{fig:aid_overview}
\end{figure*}

Type 1 Diabetes (T1D) is a metabolic disorder where an individual's
pancreas stops producing insulin. To compensate, they inject
synthetic insulin. Mobile applications, called automated insulin
delivery systems, use subcutaneous sensors to continuously monitor
glucose concentrations (glucose is the body's primary energy source)
and regulate glucose by automatically injecting insulin via an insulin
pump (Figure \ref{fig:aid_overview}). However, insulin is a dangerous
hormone as too much insulin can kill people in a matter of hours
\cite{cryer2012severe} and too little insulin can kill people in a
matter of days \cite{benoit2018trends}. For automated insulin delivery
systems, the key challenge is to maintain a balance: provide enough
insulin to prevent dangerously high glucose levels while avoiding
excessive insulin that could lead to life-threateningly low glucose
levels. In this context, security means ensuring integrity for insulin
dosing against malicious attacks or inadvertent errors.

Several commercial \cite{beta_bionics, tandem_iq, omnipod} and
open-source \cite{loop, openaps} automated insulin delivery systems
exist today, but none of them consider security as a primary design
constraint. The complexity of these systems makes it difficult to
reason about their correctness, which is crucial for security. For
example, the OpenAPS \cite{openaps} core function for calculating
insulin doses consists of 1192 lines of Javascript code, 63 input
and configuration parameters, and 90 branch statements. Furthermore,
automated insulin delivery systems that use the OpenAPS algorithm
embed it into native apps using a WebView for Javascript interpretation,
increasing the overall complexity and attack surface. Vulnerabilities or
attacks on such systems can be severe and fatal.

In this paper, we take on the challenge of building the first
trustworthy automated insulin delivery system, called \system.
Our design includes: (1) an architecture where we apply separation
principles for isolated and simple components, (2) a novel security
mechanism and policy to enable secure insulin delivery, (3) a mechanism
to account for unprecedented changes in human physiology, and (4)
the application of formal methods to prove correct the implementation
of our insulin delivery path, the most sensitive part of the system.
Despite these efforts, we identify scenarios that require users to
take action for security. Thus, we incorporate humans as part of our
defensive strategy, providing them with agency and trust to take
the required, corrective actions. In contrast to existing systems,
\system enables the use of any algorithm (including ML) to calculate
insulin doses, while ensuring security by moderating insulin doses
using 25 lines of formally verified Swift code.

Our approach is distinguished by its holistic consideration of the
entire problem of automated insulin delivery. We provide clear
security boundaries, mechanisms and policies to mitigate attacks on
vulnerable components, and formal methods to provide assurances that
our implementation is correct. Unlike prior research that exclusively
focuses on simulation for evaluation \cite{zhou2021data, zhou2023knowsafe},
we also evaluate our approach with an end-to-end system used by
individuals to manage diabetes with all the uncertainties of the
real world.

Given their life-threatening stakes, automated insulin delivery systems
require special consideration beyond the traditional purview of the
systems and security communities. Research on operating systems or web
browsers can list scores of CVEs from current systems and show how their
system improves security by eliminating their impact. Our position is
that we need to make sure that automated insulin delivery systems are
never assigned CVEs, because vulnerabilities can kill people. Thus, we
focus on building a new system, using security-first principles, to
avoid security vulnerabilities and mitigate their damage from the
start.

Our main contribution lies in a new end-to-end system that we built
and the lessons learned from our experience designing, implementing,
and deploying \system to humans to manage their T1D.  Our novel
contributions include:
\begin{itemize}
  \item Security mechanisms and policies that protect individuals from
    malicious dosing algorithms, as well as drastic changes in human
    physiology.
  \item A case study describing our experiences from a real-world
    deployment with seven people, and results from simulation to show
    that our techniques generalize.
  \item A human-centric design that considers the user as an integral
    part of the system, influencing both the architecture and
    including them in our defensive strategy.
\end{itemize}

We give top priority to the health, safety, and ethics of the humans
using \system to manage their T1D. Section \ref{sec:safety_summary}
summarizes our ethical considerations.

There are 8.4 million people living with T1D worldwide
\cite{t1d_numbers}, underscoring the need for providing trustworthy
systems for automated insulin delivery. Our cross-disciplinary
work, carried out with rigorous medical and ethical oversight, is a
first step in addressing this need.

\section{Overview}
Doctors and healthcare professional recognize Type-1 Diabetes as a
predominantly self-managed condition and support the adoption of
automated insulin delivery systems for management
\cite{pho, khor2023individualized, midyett2023one}.

This paper describes our design for \system, a system for trustworthy
automated insulin delivery; we have two primary goals.  First, we want
the software to be simple and correct. Second, we want to identify the
most vulnerable parts of the system and design a system to prevent
attacks on these components or withstand successful attacks.  This
section describes our overall architecture for \system that strives to
achieve these goals, and our threat model.

\subsection{Locking down the insulin delivery path}
Our system architecture focuses on providing security mechanisms and
policies on the insulin delivery path. The insulin delivery path in
automated insulin delivery systems includes models for calculating the
amount of insulin still active in the individual (insulin can take up
to six hours for complete absorption), closed-loop algorithms for
calculating and delivering insulin doses that could use deep neural
networks to predict future metabolic states. Together, these
components are both the most complex, and the most sensitive from a
security perspective because they are responsible for insulin delivery. 

\subsection{Architecture}
From a system design perspective, we strive to keep our implementation
simple and enable our use of formal methods to prove correct the most
critical parts. For our software system architecture, we use
separation principles from the OS and microkernel areas
\cite{accetta1986mach, hartig1997performance} applied to the
application layer to push complexity away from the most critical parts
of the system. This architecture is similar to secure web browsers
\cite{grier2008secure, wang2009multi, tang2010trust, reis2019site},
which also apply OS separation principles at the application
layer. Figure \ref{fig:biokernel2} shows how we decompose our
automated insulin delivery system, where we strive to keep our trusted
computing base simple. \system includes a \kernel that forms our core
trusted computing base. The \kernel logs events that other components
can ingest to learn the state of the system. These other components
include apps for training ML models, visualizing statistics, an AI
agent for suggesting dosing during meals, and an app for replaying
closed-loop algorithm execution. Together, these components make up
the overall \system system.

\begin{figure}[t]
\centering
\includegraphics[width=\columnwidth]{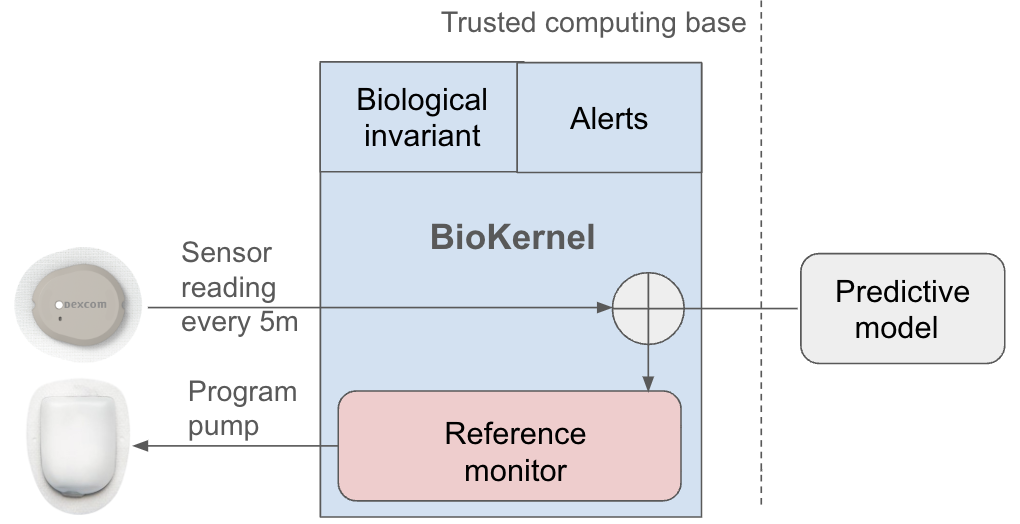}
\caption{Overall architecture for the \kernel.}
\label{fig:architecture}
\hrulefill
\end{figure}

Figure \ref{fig:architecture} shows our architecture for the \kernel.
The \kernel is the component that interacts with the continuous
glucose monitor (CGM) and insulin pump hardware, executes the
closed-loop dosing algorithm, runs security and safety checks, and
produces event logs that other components use to learn the state of
the system.

The \kernel handles the typical operations of automated insulin delivery
systems. It runs the closed-loop algorithm every five minutes as the
new CGM readings come in. Once the \kernel receives new CGM readings,
it collects the current metabolic state of the individual and sends it
to a predictive model. The predictive model calculates the amount of
insulin to inject based on its assessment of the individual's current
metabolic state and predictions of future metabolic states. The
\kernel then issues commands to the pump to deliver the calculated
amount of insulin.  A pump driver converts these commands into
low-level I/O to program the pump and adjust insulin delivery.

To withstand attacks, the \kernel serves as a \emph{reference monitor}
\cite{anderson1972computer} to enforce security policies on the
predictive model's pump commands. This reference monitor architecture
ensures that we can withstand attacks from malicious predictive models
without having to know the internals of how they operate. Our reference
monitor enforces algorithmic security, where we bound the amount of
insulin injscted by the model to stay within dynamic safety bounds
computed by a simple \emph{reactive safe model}.

Our use of formal methods focuses on the software in the insulin delivery
path. We use Hoare logic to prove key functions correct, define system
states and transitions to handle runtime verification check failures
safely, and we introduce the notion of a novel \emph{biological
invariant.} Our biological invariant is an end-to-end check of our
theoretical calculations of how the individual's metabolism should behave
compared to what we observe in practice. Violations of this invariant
suggest the need to temporarily disable automated insulin delivery in
response to an unprecedented change to human physiology.

\subsection{Threat model}
\label{sec:threat_model}
Automated insulin delivery systems inject synthetic insulin to regulate
glucose concentration in the body. In the long run, high glucose or
hyperglycemia leads to dangerous health outcomes such as kidney failure,
heart disease, amputation etc. The most serious and immediate
health risk for people who use synthetic insulin is low glucose, or
hypoglycemia. Hypoglycemia results from an overdose of insulin. Once an
automated insulin delivery system has delivered
insulin, there is no way to remove it from the body. If untreated,
severe hypoglycemia can lead to loss of consciousness, seizure, or
death in a matter of hours. Thus, our primary security objective is to
reduce or eliminate hypoglycemia.

Since personalization is important for T1D self-management \cite{glucomaker},
our design philosophy centers around humans. We provide flexibility
for people to use any model for insulin dosing that suits their preferences.
Our goal is to make sure that they stay safe even when they pick vulnerable
or malicious models. Attackers may target automated insulin delivery systems
to directly harm the human using the system.

For safety, we focus on protecting users from incorrect insulin doses
that emerge from both, poorly chosen insulin dosing models, as well as
inadvertent mistakes from benign models. Malicious or vulnerable models,
if unchecked, can be devastating because they directly affect insulin
delivery and can lead to life-threatening hypoglycemia.

We build \system as a complete iOS app, and accordingly assume that the
Swift type system, that forms the backbone of the iOS ecosystem, is
secure. We only consider FDA-approved CGMs and pumps. With the FDA being
stringent with security guidelines for medical devices \cite{fda_medical},
we assume that CGM and pump hardware are free of attacks. However, we
assume that CGMs and pumps are prone to measurement errors within their
standard error bounds \cite{garg2022accuracy, dexcom_error, Zisser2011OmniPodAccuracy}.

We discuss malicious pump drivers and ways to overcome them in
Appendix \ref{sec:driver_security}.

\section{Design principles}
Four key principles guide our design of \system:

\begin{itemize}
\item{\textit{Principle 1: Focus on simplicity for security mechanisms
    by using domain knowledge.}} Rather than design a general security
  mechanism for a broad class of problems, we focus specifically on the
  highly-risky domain of automated insulin delivery. Using domain-specific
  insights, we provide simple security mechanisms that both humans and
  theorem proving software can easily reason through.
\item{\textit{Principle 2: Design the system to empower humans to
    provide security.}} Automated insulin delivery represents an
  extreme example of high-stakes healthcare. For people to be able to
  use a secure system to manage their health, they need to have agency over
  the security of their system. Rather than simply ``alerting the human'' if
  something unexpected happens, we empower the humans using \system to be
  active participants in the security of their system.
\item{\textit{Principle 3: Consider asymmetries in risk.}} Insulin
  dosing possesses an inherent asymmetry where overdosing poses
  greater immediate danger than underdosing \cite{cryer2012severe,
    benoit2018trends}. With \system, we optimize our mechanisms to be
  more aggressive when protecting individuals from insulin
  overdose.
\item{\textit{Principle 4: Adaptability for security.}} Although we
  carefully design our software to be correct, humans and glucose
  metabolism can change quickly. Any change that affects the correctness
  of models of glucose metabolism would compromise the security of the
  entire system. Thus, we design \system to adapt to
  constantly changing human physiology to provide security both in our
  formulation of the problem and our use of formal methods.
\end{itemize}

Although we do strive to apply these principles to the entire design,
there are times when there is tension between them. For example, our
biological invariant (Section \ref{sec:bioinvariant}) helps with
adaptability (Principle 4) by detecting when our physiological models
are wrong and considers asymmetries in risk (Principle 3). But to
ensure that the human remains empowered (Principle 2), we use
necessarily complex logic to deal with cases when the biological
invariant does \emph{not} hold (violation of Principle 1). These types
of tensions are fundamental to a real-world system that manages human
health and underscores the importance of building real systems to gain
experience to navigate these tradeoffs.

\section{Algorithmic security: Withstanding malicious dosing algorithms}
\label{sec:security_design}
We have dual goals for our algorithmic security mechanism of being
both defensive and enabling. Our security mechanism is defensive by
protecting people from inappropriate insulin doses emerging from
bugs, corner cases, or even attacks merging from complex predictive
algorithms -- be it deep neural networks or massive sets of hand-crafted
rules. Our algorithmic security mechanism is enabling in allowing people
to employ any predictive algorithm to personalize the system to suit
their individual preferences.

To enable predictive algorithms, \system's algorithmic security
mechanism introduces a novel architecture that separates the use of
predictive models for insulin dosing from accounting and safety logic
to ensure that they are operating safely. This architecture provides
flexibility in using any predictive model while enforcing bounds on
insulin doses with a simple and formally verified model for safety.

\subsection{Need for algorithmic security}
Closed-loop automated insulin delivery systems have traditionally used
reactive physiological models (such as PID controllers \cite{weinzimer2008fully,
  kushner2018data}) to calculate insulin doses. These models essentially use
standard physiological models to react to offset the amount of excess glucose
in the body. This makes it easy to reason about the correctness of their
insulin doses. However, because synthetic insulin is slow-acting and takes
up to six hours to fully absorb, the reactive nature of these models limits
their effectiveness in regulating elevated glucose concentrations. Thus,
having the ability to inject insulin proactively by predicting glucose
concentrations leads to more effective glucose regulation. But, predictive
dosing opens up the individual to the risk of potentially dosing too much
insulin, and current insulin delivery systems have been reluctant to adopt
the most advanced forms of predictive models.

Researchers have shown that modern advances in ML, such as
deep neural networks, work well in making predictions for automated
insulin delivery \cite{Zhu2022, Jacobs, Zhang2021, Josep2020, RL, Tyler2022}.
Despite their promise, none of the existing automated insulin delivery
systems use deep neural networks. Instead, they use statistical methods,
like linear regression or model predictive control \cite{physio2011,
  physio2019, lunze2013control, control2019, openaps}. The reason for
using this more traditional form of ML is sound. It provides consistent,
explainable results and has a physiological basis that people can use
to vet its decisions. While advanced deep neural networks have more
predictive power, current systems do not incorporate them due to their
black-box nature and the potentially dire consequences of mispredictions.
Even a well-tested ML model does not guarantee immunity to misprediction
\cite{nfl_ml, nfl_adversarial}, especially when predicting human metabolism
that is influenced by numerous ever-changing factors such as food, exercise,
stress, environment, allergies, puberty, temperature and so on \cite{fourtytwofactors}.
An automated insulin delivery system's unchecked misprediction or
``hallucination'' \cite{chatgpt_hallucination} could have lethal consequences
if it delivers an inappropriate insulin dose \cite{cryer2012severe, benoit2018trends}.

Beyond the use of ML, the complexity of rule-based predictive models (such
OpenAPS' thousands of lines of code \cite{openaps}), make it difficult to
reason about their correctness, even if they are generally
effective at regulating glucose for diabetes management. Additionally,
people requesting enhancements to such models \cite{cob, hypo_protect}
based on their personal struggles further exacerbates model complexity.

Thus, avoiding predictive models for insulin delivery comes at the cost of
effective diabetes management. On the other hand, incorporating them
introduces safety concerns stemming from challenges in the explainability
of complex code and deep neural networks.

\subsection{Mechanism fundamentals}
The key novel insight underpinning our security mechanism design is
that all correct insulin dosing algorithms, regardless of whether they
are predictive or reactive, will deliver the same amount of insulin
over a sufficiently long period (such as a full digestion-absorption
cycle). People inject insulin to facilitate glucose absorption after
eating. Digestion converts food to glucose, and insulin facilitates
the absorption of glucose by tissues (e.g., the brain) where the body
can use it as an energy source. Since insulin dosing is dominated by
food consumption, all correct algorithms should dose the same amount
of insulin to cover glucose from digestion. But the timing of these
insulin doses has a large impact on the individual's glucose
regulation.

Our algorithmic security mechanism pairs two models. Our first model
is a predictive model that anticipates future metabolic states and
doses insulin proactively. Our second model is a reactive safe model
that measures the current metabolic state and doses insulin based only
on what it can measure currently. This pairing balances proactive
insulin dosing for faster responses with a more conservative approach
for safety. As per our insight, the reactive model would dose enough
insulin eventually, so we use it as our baseline but provide
predictive models with enough buffer to inject insulin earlier and
safely.

Architecturally, in a novel approach for automated insulin delivery
systems, the predictive model runs outside of our trusted computing
base while the reactive safe model is part of the trusted computing
base. This separation enables individuals to use whatever predictive
models they want, while maintaining security. It also defends against
malicious predictive models, as all insulin delivery decisions are
ultimately vetted by the reactive safe model. This design also enables
people to update their predictive models without violating core safety
guarantees. 

Conceptually, the predictive model is the primary model in
our system and should control the pump most of the time. We use the
reactive safe model for accounting so that we can track how far the
predictive model deviates from a known-to-be-safe baseline. Then, we
bound the size of this deviation. By combining these two models, we
strive to get beneficial properties from both. Our approach combines
the safety of a reactive model and the performance from a predictive
model.

Our security mechanism draws inspiration from previous work on
combining simple and complex models for safety \cite{sha2001using,
  mohan2013s3a}. However, our approach extends these ideas by
providing a novel security mechanism specifically designed for insulin
delivery models.  Our key contributions lie in our insight into the
security properties of insulin delivery algorithms and our principles
for how to apply our mechanism to automated insulin delivery systems
(Section \ref{sec:model_principles}).

\begin{figure}[t]
\centering
\includegraphics[width=0.75\columnwidth]{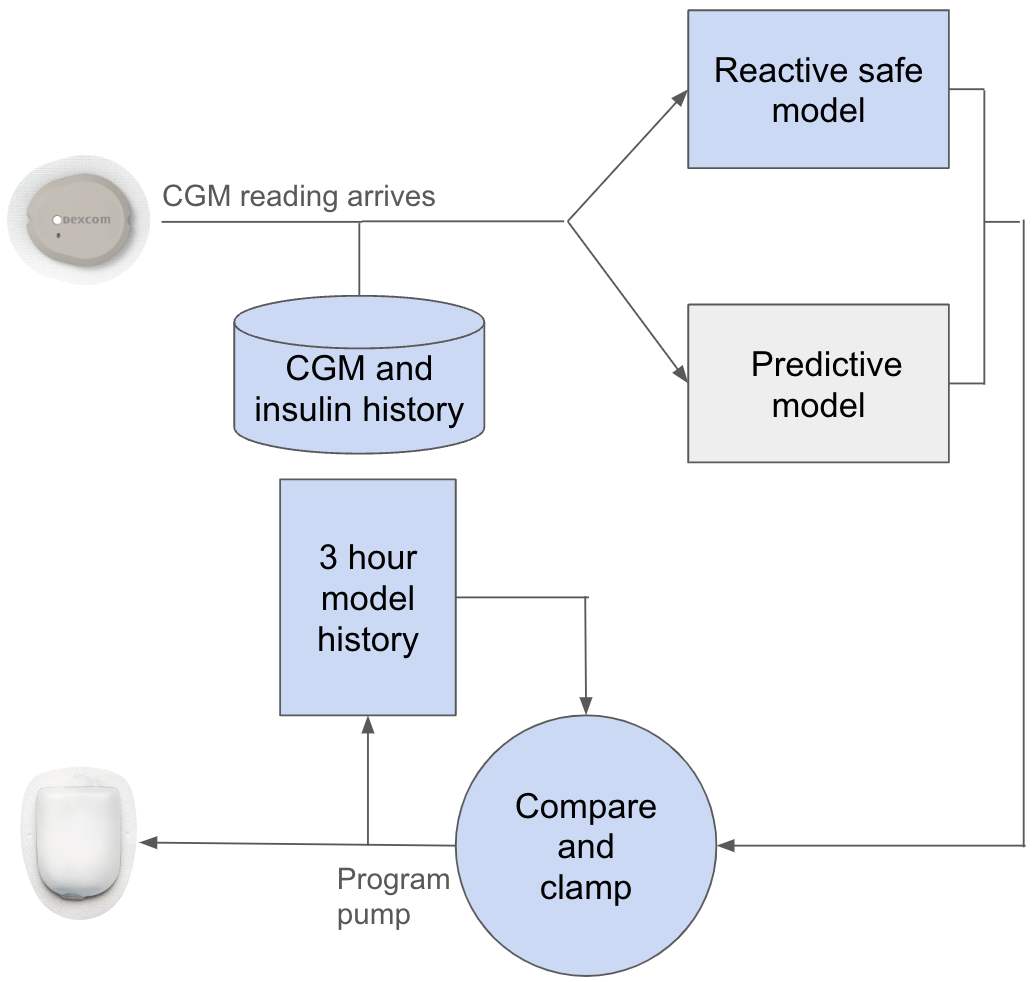}
\caption{Algorithmic security mechanism for \system.}
\label{fig:clamp_mechanism}
\hrulefill
\end{figure}

Figure \ref{fig:clamp_mechanism} shows how our algorithmic security
mechanism works. The whole process starts when a new CGM reading
arrives. Once the \kernel receives a new CGM reading, it sends the
recent history of CGM readings and insulin doses  to both the
reactive safe model and the predictive model. These models independently
produce pump commands (called a temporary basal command) that set
the insulin delivery rate for the pump over the next 30 minutes.
The \kernel then collects both commands and looks back at the differences
in doses between both models over the past three hours to calculate
how much insulin the predictive model has dosed. If the predictive
model is still within the safe insulin delivery bounds relative to
the reactive safe model's doses, the \kernel uses the predictive
model's command to program the pump. If the predictive model has
exhausted its limits, we clamp its calculated insulin to fit within
the insulin bounds relative to the reactive safe model.

To set bounds, we base our calculations on human physiology. People
produce a background level of glucose for energy. 
Our method for setting bounds allows the predictive
model to use up to three hours worth of background insulin in advance.
If the prediction is correct and glucose levels rise (e.g., from eating),
the proactive insulin will manage it. If the prediction is wrong and no
extra glucose appears, the body's ongoing glucose production will use
the extra insulin over the next three hours.

\subsection{Model principles}
\label{sec:model_principles}
The mechanism described above can be implemented with various algorithms.
Our contribution is \emph{not} in the specific algorithms we use, but
rather in the insight around insulin delivery being equal for all
correct algorithms and how we use this insight for security. In this
section, we outline the algorithms we use in our current
implementation and define the principles for how to apply security to
any predictive insulin dosing algorithms.

Algorithmically, we use a PID controller \cite{franklin2002feedback}
from feedback control for our reactive safe model, and we use a deep
neural network from the literature for meal predictions for our
predictive ML model \cite{mosquera2023enabling}. Although we use this
model in our implementation, our system supports any predictive model.
The design of the reactive safe model is crucial for algorithmic security,
as it serves as the anchor for the security mechanism. The ideal reactive
safe model should be safe, easy to understand, and have a formally verified
implementation (Section \ref{sec:formal_verification}). To ensure the safety
of this model, we base its calculations solely on facts observed automatically
from the CGM and the insulin pump. Given these high-quality sources of data,
and a simple insulin absorption physiological model, we can calculate how much
insulin an individual needs and how much previously injected insulin has not
yet taken effect. From this calculation, we know how much insulin to inject or
withhold.

These calculations are standard calculations that people who inject
insulin manually use to determine dosing, making them easy for people
who manage T1D or their medical care team to understand. The
difference in our system is that we run these calculations
automatically and every five minutes to adjust to the latest sensor
readings. This basic formulation is standard and used in most
automated insulin delivery systems.

Our evaluation shows that a benign predictive ML model outperforms the
reactive safe model in terms of glucose regulation, both in simulation
(Section \ref{sec:eval}) and for the individual who uses our system
for their real-world insulin dosing (Section \ref{sec:case_study}).
Our evaluations in simulation also demonstrate the fatal consequences
of using an incorrect or malicious ML for insulin dosing, as well as
our mechanism's ability to withstand such models to protect
individuals (Section \ref{sec:eval}).

\section{Holistic security: Biological invariant}
\label{sec:bioinvariant}
We recognize that human physiology is complex and it is difficult to
ensure that the \kernel's reactive safe model, which forms the basis
of algorithmic security, can handle all changes to human
physiology. Concretely, since the reactive safe model incorporates
parameters that represent the user's physiological state, any
divergence in their values could render the model ineffective in
protecting the user. With \system, we introduce a \emph{biological
invariant} to detect when such divergence could endanger the user, and
shut off automatic insulin dosing to protect them.

From a high level, the biological invariant runs experiments every 5
minutes to empirically measure physiological parameters.  These
experiments compare the expected drop in glucose from prior insulin
doses against the observed change in glucose sensor readings, thereby
providing an end-to-end security mechanism. The biological invariant
effectively serves as a ``catch-all'' off-switch for insulin delivery
when the \kernel's view of the user's physiology becomes
inaccurate. Although our reactive safe model adapts to small changes
in physiological parameters by design, the biological invariant
detects large changes that operate outside of the bounds of our
design.

Insulin sensitivity, which is the amount of glucose absorbed per unit
of insulin, is a key physiological parameter that can change. Changes
to insulin sensitivity happen from drastic changes to the user's
glucose absorption, such as during exercise, when the body uses
glucose without needing insulin \cite{borghouts2000exercise}, or if
the pump insertion site hits a vein and delivers insulin directly into
the blood stream \cite{steyn2023accelerated}. In such cases, the user
becomes more sensitive to insulin where the same amount of insulin
will facilitate more absorption of glucose than expected, increasing
the risk of insulin overdose. For example, if the user becomes twice
as sensitive to insulin while jogging, the same insulin dose would
decrease their glucose by twice the expected amount, pushing them
towards dangerously low glucose (hypoglycemia).

To implement the biological invariant, we compare the actual change in
the user’s glucose, against the \kernel's calculations of how much
their glucose \emph{should} have changed based on current insulin doses
over a fixed time window. If the observed drop in glucose deviates beyond
a clinically determined threshold of 30 mg/dl \cite{30mg_1, 30mg_2} from
the calculated drop in glucose within a window of 30 minutes, we consider
the biological invariant to have been violated and shut off insulin delivery.
Mathematically, we define the biological invariant as follows:
\begin{equation}
  \Delta G^{C} - \Delta G^{A} \leq T \tag{1}
\end{equation}

Here, $\Delta G^{C}$ denotes the \kernel's calculated drop in glucose based
on insulin doses over a 30 minute window, $\Delta G^{A}$ denotes the actual
drop in the user's glucose over the same time window, and $T$ is the threshold
of 30 mg/dl. Crucially, we note that we cannot precisely calculate
$\Delta G^{A}$ or $\Delta G^{C}$ since CGMs and pumps are prone to measurement
errors. To account for these errors under standard use, we enforce error bounds
on readings from CGMs \cite{garg2022accuracy, dexcom_error} and 
records from pumps \cite{Zisser2011OmniPodAccuracy} under standard conditions.
Based on these bounds, we check the following condition on the \kernel's current
estimate of insulin sensitivity:
\begin{equation}
S \leq \frac{0.9 \cdot G_{t+\Delta t}^{M} - 1.1 \cdot G_{t}^{M} + T} {B \cdot \Delta t - 0.99 \cdot \Delta IOB^{M} + 0.01 \cdot \Sigma I_{i}^{\
    M}} \tag{2}
\end{equation}

Here, $S$ is the \kernel's estimate of the user's insulin sensitivity,
$G_{t}^{M}$ and $G_{t+\Delta t}^{M}$ are readings from the CGM, $\Delta IOB^{M}$
and $\Sigma I_{i}^{M}$ are calculated from the insulin doses recorded by the pump,
$\Delta t$ denotes the 30 minute window, and $B$ is the user's basal rate, a physiological
parameter that often remains stable for days and is less prone to change than insulin
sensitivity \cite{Lindmeyer2021BasalProfiles}. We refer to this condition as the
\emph{implementation invariant} and use formal methods to prove that violations to
this condition imply violations to the biological invariant (Section
\ref{sec:formal_verification}). Since all quantities in the implementational
invariant are observable, we shut off insulin delivery when it is violated.
We show how to adopt mathematical models from the literature
\cite{simglucose, physio2019} to derive the implementational invariant in
Appendix \ref{sec:invariant_derivation}.

We resume normal execution if the implementational invariant goes back
to being satisfied within two hours, based on clinical guidance for
exercise for people living with T1D \cite{ucla_t1d, borghouts2000exercise}.
If it remains violated for more than two hours, we alert the user and
transition to manual insulin delivery.

Appendix \ref{sec:low_sensitivity} discusses the less dangerous scenario
when the actual change in glucose is higher than the calculated change.

\section{Last line of defense: Humans}
\label{sec:humans}

In this section, we describe our design and implementation for
incorporating humans as a critical component of our multi-layered
security system. While our algorithmic security mechanism and biological
invariant defenses form the primary barriers against attacks, humans
serve as the last line of defense, to take corrective action against
hypoglycemia. We discuss our most interesting findings around
user preferences and the architectural iterations to our design from
real-world experiences.

\subsection{Predictive alerts for humans}
When an individual's glucose levels are not within the range of healthy
values, they need to intervene to bring them back in range. Low glucose,
or hypoglycemia, occurs from too much insulin, and there is no way to
remove insulin that the automated insulin delivery system has already
delivered. To recover from hypoglycemia, people need to eat sugar.
Given that people with T1D are already active participants in managing
diabetes, we focus on how we can use them to create a more comprehensive
security system.

\begin{figure}[t]
\centering
\includegraphics[width=0.65\columnwidth]{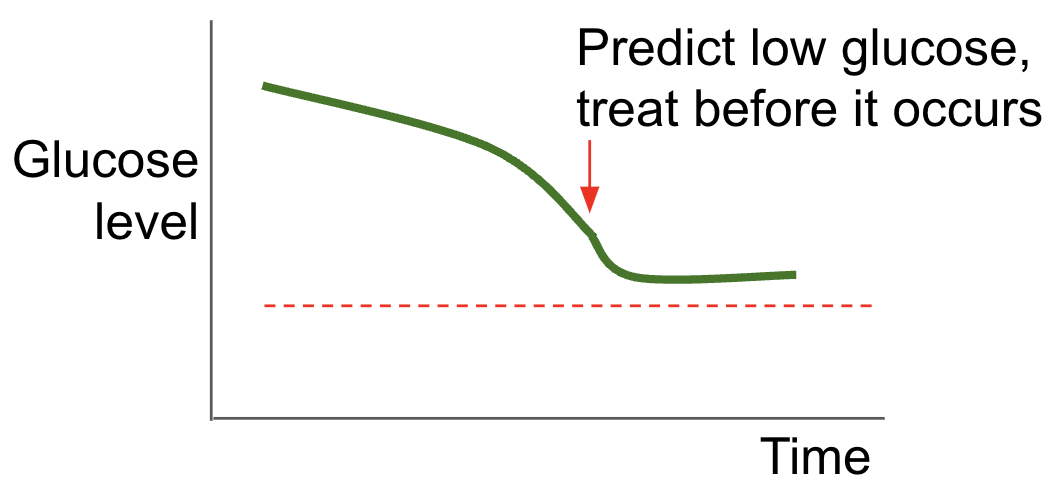}
\caption{\system predicts hypoglycemia and notifies the user
  before it happens so they can proactively prevent it.}
\label{fig:low_glucose}
\hrulefill
\end{figure}

We use predictive yet deterministic mobile notifications to alert
people about impending hypoglycemia before it happens (Figure
\ref{fig:low_glucose}). We focus on hypoglycemia since it presents
the most immediate and dangerous risk. Alerts for hypoglycemia ae not new.
All CGM software comes with non-predictive hypoglycemia alerts by default
\cite{dexcom_alerts} and glucose prediction algorithms have been well
covered in the literature \cite{Josep2020, Zhang2021,
  arora2023multivariate}. Our contribution is using predictive alerts
before someone reaches hypoglycemia as a security defense.

In a user study (Section \ref{sec:predictive_alerts_eval}), we reduced
the amount of time six people spent in hypoglycemia as a result of
using \system, despite them already having CGM alerts set up prior to
our study. We highlight that we did not conduct our user study in a
controlled environment, but had participants use our app for predictive
alerts as they managed their diabetes in the real world. We also
highlight that The number of participants in our study is also roughly
consistent with recent human factors in computing research
\cite{glucomaker, moont1d, context_aware_t1d, t1d_children}. Our results
show that predictive alerts can be effective as a security measure
against hypoglycemia.

Using a simple and deterministic algorithm is important since it runs
as a part of our trusted computing base. Algorithmically, we predict
hypoglycemia using linear regression over the last 20 minutes of
CGM readings and extrapolate forward 15 minutes. We trigger alerts if
our prediction crossed the hypoglycemia threshold in the next 15
minutes.

The most interesting finding from our study was that personalization
was important. We provided people with the ability to set the
prediction threshold and specify how often they wanted the alert to
repeat while they remained in hypoglycemia.  Five of the six participants
customized their alerts and the remaining participant noted that the
default values we picked coincided with their alerting preferences. In our
first implementation we did not provide the ability to customize their
settings and we tried to make the usability vs security tradeoff
algorithmically and automatically. But in the end, simplifying the
system to enable the personalization of alerting policy was important
for our participants.

Our finding has the potential to have broader implications for
security in healthcare systems. Often, security systems favor
uniformity and consistency \cite{DHILLON2016656, FURNELL20165}, but
when managing people's health we found that allowing people to balance
alert fatigue and security increased engagement and potentially
increased protection, based on each individual's needs.

\subsection{Architectural implications}
One interesting design iteration we made was moving the alerting logic
into the \kernel itself. In our first design, consistent with our
separation principles, the alerting logic resided in a separate
app. This alerting app learned the state of the system using our
cloud-based event logging system. However, one of our participants
went on a canoeing trip and was disconnected from the internet, so
they could not get alerts. Ideally iOS would provide inter-process
communication mechanisms to pass data from the \kernel directly to the
alerting app, but iOS does not have the appropriate APIs for this
style of communication. Thus, we moved the alerting logic into the
\kernel, violating our principle of separation, but enabling our users
to get alerts even when they are disconnected from the internet.

Moving the alerting logic into the \kernel did have security
implications. It adds complexity to our trusted computing base,
but we carefully manage this tradeoff. We minimize the added
complexity by using traditional, well-understood algorithms for
prediction rather than complex ML models and reuse existing
functionality within the \kernel when possible to keep the
implementation simple. This approach allowed us to maintain a high
level of confidence in the security and correctness of our alerting
system while still providing the necessary functionality.

\section{Formal verification}
\label{sec:formal_verification}
We formally verify the most critical components of
\system to prevent unexpected and undesired outcomes
\cite{Thomas1994TheSO}. Specifically, we verify the
implementation for: (1) delivering bolus insulin doses (a single dose
delivered immediately), (2) delivering basal insulin doses (multiple
doses delivered at a constant rate over a specified time period), (3)
clamping insulin doses using the reactive safe model, and (4) enforcing
the biological invariant. We consider these components to be the most
critical since they implement our security mechanisms and directly
control insulin delivery.

Formally verifying our \system's implementation in Swift presented
an interesting challenge since automatic translators do not exist
between Swift and languages for verification, such as Dafny. While
automatic translators exist between Dafny and other languages (such
as C) we could not pick those languages over Swift for implementation.
This is because Swify provides the most support to build iOS applications
and our goal was to build a complete automated insulin delivery application
for real-world use.

To overcome this limitation, we manually ported our Swift implementation
to Dafny. While manual translation is prone to human error, we took
important steps to reduce the likelihood of such errors. Concretely,
we ported our Swift code line-by-line to Dafny, preserved variable and
function names across implementations \cite{chong2020code}, and wrote
unit tests for each function in both languages to ensure behavioral
consistency. 

We establish invariants among the aforementioned critical
components along the insulin delivery path \cite{Thomas1994TheSO}.
These invariants enforce critical constraints such as maintaining
appropriate units insulin doses, disallowing immediately
successive bolus doses, ensuring basal doses match doses
supported by the pump etc. Collectively these invariants are
important to ensure the expected safe operation of the \kernel.
We discover 9 bugs in our original implementation by verifying
31 invariants across 8 functions. Other than the biological invariant
(Section \ref{sec:bioinvariant}), some of our other invariants enforced
that there should be a gap of at least 4.2 minutes between bolus doses,
the amount of insulin on board should also take into account the
insulin from basal doses etc. We list all our invariants
and bugs in Appendix \ref{sec:invariants_list}.

From our experience with formal verification, we report two key
takeaways about our implementation for secure automated insulin
delivery. First, we were able to prove the correctness of even
those components that involve unobservable quantities (such as
the user's true physiological state in the biological invariant)
under standard operation. Second, while we expected most bugs
to reside within the implementation of core algorithms, they
were fully concentrated among peripheral components
such as user settings.

\subsection{Capturing true physiological state}
Glucose readings from CGMs and insulin doses recorded from pumps
do not reflect the user's true physiological state. This is
because CGMs and pumps are subject to measurement errors, making
precise glucose concentrations and insulin doses unavailable
in Swift.

To model this discrepancy, we use ghost variables in Dafny
\cite{Dafny} to represent the true glucose values and insulin
doses. We then introduce \emph{device invariants} that bound
the maximum measurement error within standard use, i.e., 10\%
for CGMs \cite{garg2022accuracy, dexcom_error}
and 1\% for pumps \cite{Zisser2011OmniPodAccuracy}. We
enforce these bounds as preconditions when verifying the
implementation of all our core components. Most importantly,
we proved the correctness of the implementational invariant
(Equation $(2)$ in Section \ref{sec:bioinvariant}) to verify
the biological invariant (Equation $(1)$ in
Section\ref{sec:bioinvariant} as a post condition in Dafny).

\subsection{Discovering bugs outside the core algorithms}
7 out of the 9 bugs we discovered with formal verification were related to
issuing insulin delivery commands with invalid parameters. With these bugs,
users could specify negative amounts of insulin for basal and bolus doses,
set 0-second duration for basal doses, allowing the time to stop a basal
dose to be preced the start time of the dose etc. Other bugs involved
insfficient error handling, for example when disallowing certain invalid
bolus doses. We did not discover any bugs in \system's core algorithms,
such as those used by the reactive safe model. We resolved bugs by updating
our Swift implementation to match the formally verified Dafny code.

\section{Implementation}
\label{sec:implementation}
In our implementation, the \kernel is an iOS app that consists of 5.1k
lines of code in our trusted computing base. It communicates with a
Dexcom G6 or G7 CGM and Omnipod DASH insulin pump via Bluetooth low
energy, and runs the core closed-loop algorithm. For the \kernel, we
use the Swift programming language, the Actor abstraction for
data-race-free concurrency, and JSON files stored on disk to capture
persistent state. Although the \kernel is a fully featured and stand
alone automated insulin delivery system, \system uses an event logging
system (Figure \ref{fig:biokernel2}) that enables separate apps to
extend functionality while remaining isolated from the \kernel.

\section{Evaluation}
\label{sec:eval}
In this section, we evaluate \system's algorithmic security
mechanism in simulation. In the next section (Section \ref{sec:case_study}),
we evaluate all components of \system in the real world. We do not
evaluate the biological invariant in simulation since existing simulators
do not support changes to human physiology.

\subsection{How effective is \system's algorithmic security mechanism?}
\label{sec:alg_eval}
We evaluate \system's algorithmic security mechanism on 21 virtual
humans from an FDA-approved simulator \cite{simglucose}. The mechanism
should protect users against untrusted models that administer incorrect
insulin doses (overdose or underdose), while minimally impeding algorithms
that are efficient at managing T1D. While subtle attacks are also possible,
we focus on extreme attacks in our evaluation since they have the most
adverse consequences.

We first run simulations with a predictive ML model inspired from prior
research \cite{mosquera2023enabling}, without incorporating our
security mechanism to set a baseline of an effctive algorithm in
managing T1D. Then, in the same simulation scenarios, we introduce an
order of magnitude increase to the baseline algorithm's insulin doses
to quantify the risks of untrusted algorithms. We then repeat these
experiments with our security mechanism to evaluate its impact.

The American Diabetes Association (ADA) recommends that individuals
with T1D spend at least 70\% of their time in the range (TIR) of
70-180 mg/dl to remain healthy \cite{nathan2014diabetes, endocrine_tir,
  journal_glycemic_targets}. Glucose levels less than 70 mg/dl (hypoglycemia)
put individuals at risk for acute complications like impaired consciousness,
seizure, and death, while levels greater than 180 mg/dl (hyperglycemia) increase
risks for long-term vascular damage. We use TIR as our metric in these
evaluations.

\subsubsection{Without algorithmic security mechanism}
\textbf{Predictive ML model baseline} With the predictive ML model, we
followed all ML best practices in carefully tuning hyperparameters and
evaluating the model on simulation scenarios that were different from
those used in training. Figure \ref{fig:det_ml_outcomes} shows the
average proportion of time spent in the recommended healthy range (TIR)
by virtual humans from different age groups when employing the predictive
ML model. We see that individuals across all groups spend over 85\% of
their time in range compared to 77\% when using only the reactive safe
model. These results provide a baseline for a predictive algorithm that
is effective in managing T1D.
  
\begin{figure}[t]
    \centering
    \small
    \begin{tabular}{|c|c|c|c|}
      \hline
           {\bf }         & {\bf Reactive} & {\bf TIR with}   & {\bf TIR with} \\
          {\bf }          & {\bf safe}     & {\bf predictive} & {\bf clamped} \\
          {\bf Age group} & {\bf model}    & {\bf ML model}   & {\bf ML model} \\
          \hline
          Adults      & 88.25\% & 93.49\% & 92.88\% \\ \hline
          Adolescents & 81.98\% & 85.23\% & 82.38\% \\ \hline
          Children    & 77.31\% & 86.6\%  & 85.42\% \\ \hline
    \end{tabular}
    \caption{Average proportion of time spent in range (TIR) across 21
      virtual humans with a reactive safe model, a predictive ML
      model, and clamping the predictive ML with the reactive safe
      model for security.}
    \label{fig:det_ml_outcomes}
    \hrulefill
\end{figure}

\textbf{Incorrect ML dosing} 
In this section, we discuss an experiment with a malicious algorithm
that intentionally doses too much insulin. Concretely, our malicious
algorithm doses ten times the amount of insulin computed by the
previously described predictive ML model. We ran this algorithm within
the simulator on the same virtual humans on the same scenarios as the
previous experiment.

\begin{figure}[t]
  \centering
  \includegraphics[width=\linewidth]{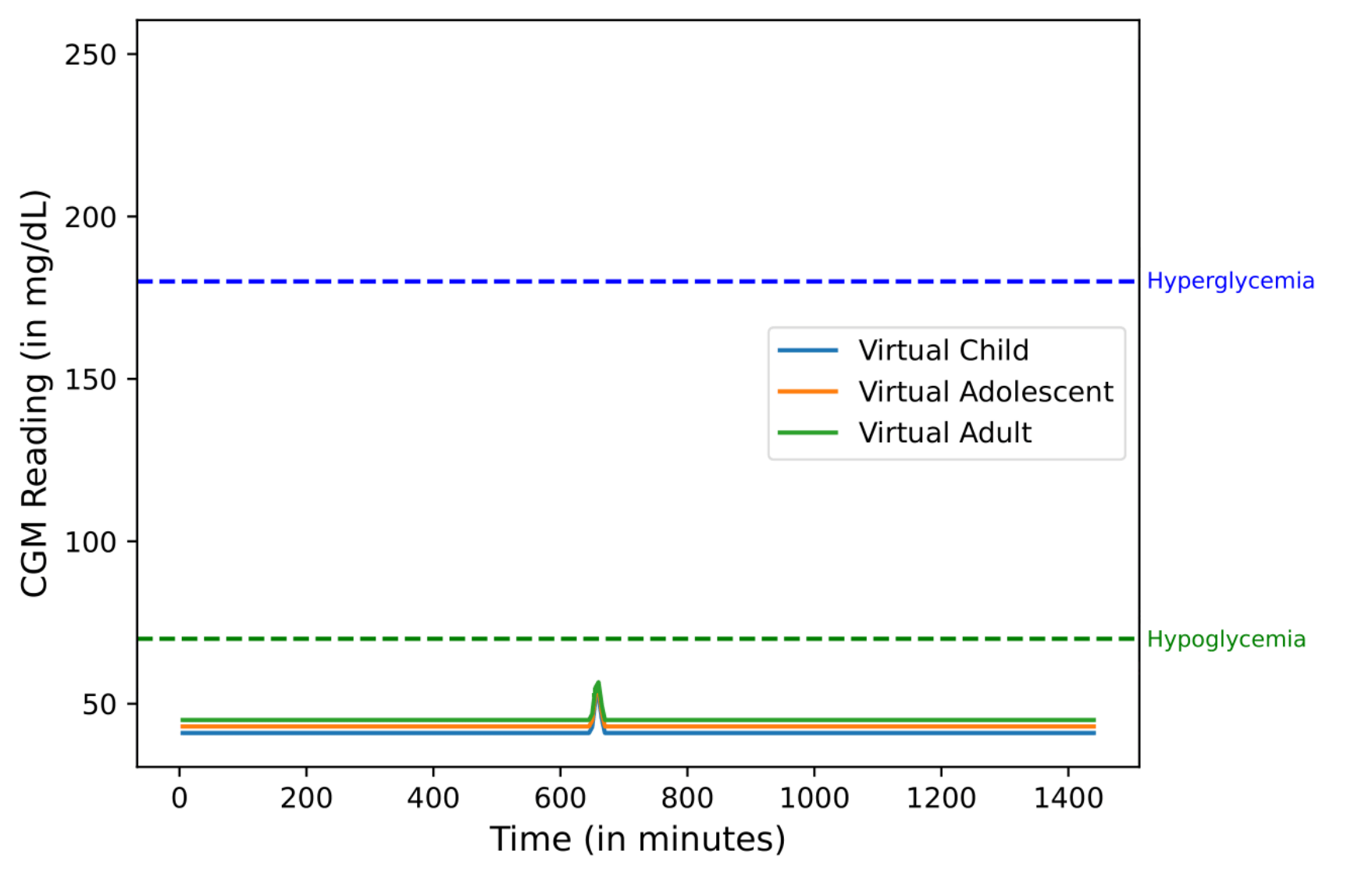}
  \caption{Variation in CGM readings for three virtual humans when
    running a predictive ML algorithm that intentionally doses ten
    times the amount of the required insulin. The CGM readings staying
    below the hypoglycemic bound would have likely been fatal to all
    three individuals.}
  \label{fig:malicious_more}
  \hrulefill
\end{figure}

Figure \ref{fig:malicious_more} shows the CGM readings on three
randomly chosen virtual humans. The readings fully go below the
hypogycemic bound, which would have killed them in the real world.
Across all virtual humans, we see an average drop in TIR from 88.44\%
to 0\% when dosing too much insulin. Correspondingly, we observe an
average increase in the time spent in hypoglycemia from 0\% to 100\%.
We report similar results with intentional underdosing in Appendix
\ref{sec:underdose_eval}. These results demonstrate the need for
algorithmic security.

\subsubsection{With algorithmic security mechanism}
\textbf{Impact on incorrect dosing.}  We repeat experiments from the previous
section with the same simulation scenarios. We run the same malicous model
to overdose insulin, but also run our reactive safe model
(Section \ref{sec:security_design}).

\begin{figure}[t]
  \centering
  \includegraphics[width=\linewidth]{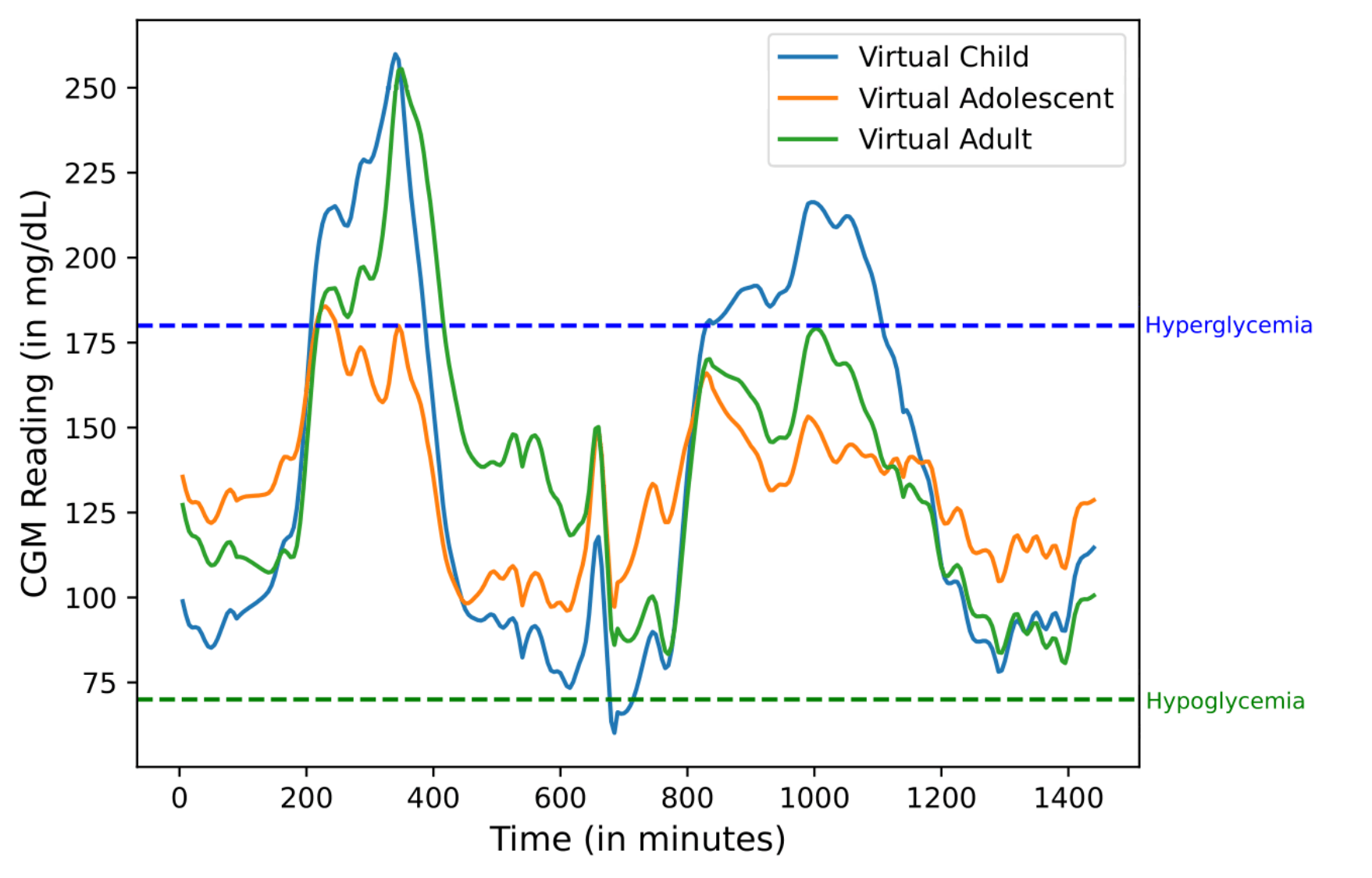}
  \caption{Variation in CGM readings for three virtual humans when clamping an algorithm that
    intentionally tries to dose ten times the amount of the required insulin. On average, CGM
    readings stay in range over 70\% of the time.}
  \label{fig:safety_more}
  \hrulefill
\end{figure}

Figure \ref{fig:safety_more} show CGM readings on the same three virtual
humans on the same scenarios as the experiment in the previous section.
We see that all of them spend over 70\% of their time in the healthy range,
showing that \system protects individuals from  malicious overdosing.
On average, across all virtual humans, we only see a drop in TIR from 88.44\%
with correct dosing to 78.22\% when employing \system while overdosing. We
also report similar results on experiments with a malicious algorithm that
underdoses insulin in Appendix \ref{sec:underdose_eval}.

While the times spent in range are less than those of correct dosing, the
impact of incorrect dosing is significantly dampened with individuals spending
the recommended amount of time in the healthy range, despite active attack. 

\textbf{Impact on effective benign algorithms.}  We evaluate the
impact of \system's algorithmic security mechanism on benign
algorithms that are effective in managing T1D such as the baseline
predictive ML model. We repeat the experiment of running the
predictive ML model on all virtual humans in the same scenarios
with the same parameters for the clamps and reactive model used to
evaluate incorrect dosing.

Comparing data in Figure \ref{fig:det_ml_outcomes} we see that there
is only a slight drop in the average proportion of time spent in range
by virtual humans across all age groups when employing \system's
algorithmic security mechanism. When employing the reactive safe model
as is to calculate insulin doses, we report an average TIR of 82.48\%
which is lower than the average TIR of 86.84\% observed when running
the predictive ML model with \system's security mechanism. This shows
that \system's algorithmic security mechanism does not impede effective
T1D management.

\section{Using \system in the real world}
\label{sec:case_study}
In this section, we report on our experiences working with seven
individuals using \system to help manage their T1D. Six people are
students who only used the predictive alerting feature of the \kernel
(Section \ref{sec:humans}) for two months, from June 2024 - August
2024. One individual, who we will call Bob, has been using the full
\system system since November 2023 for insulin delivery and predictive
alerting. The number of participants in our study is roughly consistent
with existing research in human factors in computing
\cite{glucomaker, moont1d, context_aware_t1d, t1d_children}. More importantly,
we report results on indiviuals using \system to manage their diabetes
in the real world.

\subsection{Ethical considerations}
\label{sec:safety_summary}

In this section, we outline the steps we take to ensure that we uphold
high ethical standards and ensure that the people using our system,
are safe. Our user study for the six students was approved by our
university's IRB. Bob's use \system was determined by our
university's IRB to be ``self evaluation'' and exempt from IRB.

One of the co-authors of this paper is a board-certified
Endocrinologist, who specializes in T1D. They designed the safety
protocol, which defines the specific criteria we use to stop the
study, if needed. Appendix \ref{sec:protocol} includes a more
detailed discussion of our ethics and safety protocols.

\subsection{Is an ML-based closed-loop system safe and effective in practice?}
Bob has been using \system in a closed loop since November 23rd,
2023. From November 23 to January 28th, 2024 we used our reactive safe
model to make control decisions. Starting on January 28th Bob started
to use our predictive ML model for periods of time to get used to the
new system, and then starting on January 31st, 2024 starting using our
predictive ML model exclusively. We report on one week's worth of data
from the predictive ML model starting on January 31st ending on
February 6th, 2024.

Overall, Bob had the best results when using ML. His time spent in
range increased from 96.3\% to 97.2\%, and notably he spent no time in
hypoglycemia (below 70 mg/dl), going down from 0.67\% when not using ML.
More importantly, Bob's A1C, which captures his long-term glucose
concentration, dropped to 5.8\% as a result of using \system. This result
demonstrates a significant health outcome since it matches up to
non-diabetics who have an A1C of 5.6\% or lower.

Next, we measure how often the predictive ML model was responsible for
programming the pump vs our reactive safe model. For the one week when
Bob was using the predictive ML model, 33.0\% of the time both the
predictive ML model and reactive safe model issued the same commands,
39.9\% of the time the \kernel used the predictive ML model's
commands, and 27.1\% of the time the \kernel used the reactive safe
model's commands. The reactive safe model having to take over to issue
commands is not surprising since ML models are not immune to mispredictions,
especially when taking the variability of the real world into account. We
believe that this distribution represents a suitable balance for
providing the ML with enough flexibility to improve outcomes, while having
tight enough bounds to limit potential damage from bad or malicious ML
predictions.

As of July 2025, Bob has swapped the predictive ML model with a rule-based
model similar to OpenAPS' model described in Section \ref{sec:intro}.

\subsection{Do biological invariant violations trigger in practice?}
\label{sec:biological_invariant_case_study}
To determine if biological invariant violations trigger in practice,
we review 90 days of data for Bob, from May 29th, 2024 to August 27th,
2024. During this period of time, biological invariant violations
occurred 1.6k times, or 9.0\% of the time. Of these violations there
were 49 sequences of 30 minutes or longer with the longest sequence
being 65 minutes.

These results suggest that although natural variations in glucose
metabolism do lead to biological invariant violations, none of them
were long enough to trigger our most severe action of transitioning to
manual insulin dosing after two hours. We manually reviewed one weeks
worth of Bob's data and discovered that all five violations within this
period were caused by exercise.


\subsection{Are predictive alerts useful for security?}
\label{sec:predictive_alerts_eval}
To measure predictive alerts as a security mechanism to avert
hypoglycemia (Section \ref{sec:humans}), we recruited six students
who live with T1D and had them use the \system predictive alerts to
complement their default CGM alerts. Our primary metric is time spent
below range (i.e., below 70 mg/dl), where a decrease in the amount of
time spent in hypoglycemia suggests that predictive alerts are effective
to empower people to mitigate the most immediate and dangerous security
risks.

For the two month study, our participants decreased their average time
in hypoglycemia from 2.7\% down to 1.6\%, and all individuals
experienced a decrease in the amount of time spent in
hypoglycemia. For context, participants in a state-of-the-art
automated insulin delivery clinical trial experienced a decrease of
time in hypoglycemia of 0.9\% \cite{brown2019six}, showing that our
1.1\% decrease is substantial.

We also report positive feedback from all participants during monthly
interviews. Other than customization, participants acknowledged the
importance of being able to reason about the alerts once we explained
that we used linear regression for prediction. They reported that this
helped forge trust in our system which in turn reduced their cognitive
load with diabetes. All participants have continued using our system
even after the study's conclusion.

Overall, our results show how predictive alerts are useful for automated
insulin delivery.

\subsection{How does the \kernel perform?}
We evaluate two aspects of \system's system performance in the real
world. First, we measure how long our closed-loop algorithm takes to run
and second, we count how often the closed-loop algorithm runs
successfully end-to-end. We use Bob's iPhone 14 to collect all of the
data we show and these numbers are averages across the entire eleven
month period.

On average, the \kernel spends 4.3s running the closed loop algorithm,
where almost half of that time is dominated by CGM and pump
communications. Of the components within the closed-loop algorithm,
the ML predictions take 2.2s on average and the safety logic takes
19ms (0.019s). Since the system runs in the background, we believe
that iOS runs the device with reduced computational capacity to save
power. But overall, 4.3s of wall-clock time for every five minutes our
closed-loop algorithm runs with only a tiny fraction of that coming
from our safety logic represents practical system performance.

For successful closed-loop end-to-end execution, over the evaluation
period the closed-loop algorithm runs successfully 96.5\% of the time,
with 2.1\% of the runs failing due to pump communication,
which is our largest class of failure.

\section{Related work}
The complexity of automated insulin delivery systems and the fact that
they make real-time healthcare decisions, makes their security
paramount. Current research on security for these systems focuses
primarily on evaluating algorithms using simulations
\cite{zhou2021data,zhou2023knowsafe}. Our experience shows that
when faced with the practical challenges of real systems, algorithms designed in
simulations often prove insufficient or inapplicable. These real-world
complexities require a fundamentally different approach to designing
security for automated insulin delivery systems. \system moves beyond
simulation and tackles the challenges of real-world systems.

Previous research has looked at the security of implanted medical
devices in general \cite{halperin2008security, burleson2012design,
  rushanan2014sok}, in addition to looking at insulin pumps in
particular \cite{li2011hijacking, paul2011review}, with more recent
work looking at providing improved security
\cite{marin2016feasibility, ahmad2018securing}. Also, recent work has
looked at applying formal methods to insulin pumps for high assurance
\cite{panda2021secure}. They focus on devices and their communication
channels. With \system, we assume that these devices are secure and
focus on automated insulin delivery software.


\section{Conclusion}
The design of \system, while grounded in established OS and security
principles, incorporates several non-obvious elements that emerged
from our deep engagement with the problem domain and real-world
testing.

First, our approach to algorithmic security used classic security
techniques, like reference monitors and the Simplex
architecture. However, our novel insight around all algorithms dosing
the same amount of insulin led to the application of these
classic techniques. Plus, from our exploration, the most interesting
part was \emph{not} the use ML, but the principle of repurposing
reactive models for security.

Second, the biological invariant emerged as a key check in our system,
addressing a blind spot in typical simulations that ignore changing
human dynamics. This forced us to develop novel formal methods to
handle scenarios where our glucose metabolism models could be
drastically wrong.

Third, our approach to human interaction in the system was
nuanced. We found that simply alerting humans was not sufficient.
For security alerts, personalization, customization, and simplicity
was important.

These non-obvious design decisions underscore the complexity of
creating a truly trustworthy automated insulin delivery system. They
reflect the value of combining theoretical security principles with
practical, real-world testing and domain-specific knowledge. This
interdisciplinary research drew on deep expertise from operating
systems, computer security, medicine, and ethics to create a practical
system.


\bibliographystyle{plain}
\bibliography{typezero}

\appendix
\section{Ethical and safety considerations}
\label{sec:protocol}
One individual, who we will call Bob, used \system for 9 months. Six
university students also used the predictive alerts from \system for 2
months. We put in place a protocol where we would stop our study and
ask people to stop using our system and revert back to their previous
management regime. In our protocol, we have formal in-person meetings
with Bob every week to check his data and make sure that he is not
incurring risk as a result of using \system, and weekly surveys for
the six students, plus monthly meetings. We would have stopped our
study if any of the following conditions occurred:
\begin{itemize}
  \item \emph{Excessive hypoglycemia.} We define excessive
    hypoglycemia as having spent 8\% or more of time during a week
    with CGM readings below 70 mg/dl or more than 1\% below 54
    mg/dl. These amounts of time spent in hypoglycemia would have
    represented a significant increase in time spent in hypoglycemia
    for participants.
  \item \emph{Insufficient CGM data.} If they had less than
    70\% of the time during a week with CGM readings, then the lack of
    CGM data would represent a decrease for the participants.
  \item \emph{Insufficient closed-loop runs.} We designed \system to
    run closed-loop algorithms every five minutes, or 288 times per
    day. If Bob had a day where our closed-loop algorithm ran
    successfully less than 200 times, it would represent a fundamental
    flaw in the system.
  \item \emph{Serious complications due to diabetes.} Serious
    complications include any hospitalization for diabetes related
    issues, severe hypoglycemia where a participant was unable to
    recover from hypoglycemia themselves and had to get help from
    someone else, diabetic ketoacidosis, or seizure.
  \item \emph{New medical diagnosis requiring significant attention.}
    If anyone had received a new medical diagnosis during the study,
    the study would have had the potential to be a distraction for
    participants.
  \item \emph{New mental health diagnosis.} If anyone were struggling
    with mental health during the study, the study would have the
    potential to be a distraction.
\end{itemize}

In addition to our safety protocol, we followed best practices for
ethical research to ensure that we conducted our study ethically.
\begin{itemize}
  \item \emph{IRB.} Our study went through our university's IRB
    process, where the study including six university students who
    used \system's alerting software was approved, and our case study
    on Bob was determined to be ``self evaluation'' and exempt from
    IRB.
  \item \emph{Informed consent.} All participants went through a
    written informed consent process.
  \item \emph{Data management.} \system does \emph{not} store any
    personally identifiable information and uses access control on our
    Google Cloud Platform infrastructure to ensure that only the
    authors of the paper can access this anonymized data.
  \item \emph{Risk/benefit analysis.} All participants are already
    living with T1D and use CGMs and inject insulin, so they already
    take on the risks inherent in managing T1D. The benefits include
    improved control over their T1D. There are risks in using
    experimental software, but all participants continued to rely on
    their CGM alerts as a last line of defense.
  \item \emph{Participant selection.} We selected participants who
    were university students living with T1D.
  \item \emph{Adverse event reporting.} Had any participants
    experienced adverse events during the study, we would have
    reported it to our IRB for guidance and addressed it according to
    our safety protocol.
  \item \emph{Communication.} All surveys were done through our
    university's medical school infrastructure for private
    communication with individuals.
\end{itemize}

Since Bob used the full \system software, including for
insulin delivery, we included several other safety mechanisms:
\begin{itemize}
  \item Bob increased the frequency of his visits to his
    Endocrinologist while using the \system software. He went from
    yearly appointments to quarterly appointments.
  \item During the study, twice Bob got lab work based on blood tests
    to provide an independent measure of his overall metabolic
    health. This lab work helped to ensure that he was in good health
    and that there were no hidden negative impacts from his use of
    \system.
\end{itemize}

Overall, \system had a large positive impact on the individuals who
used it. Bob's lab-based A1C test (a blood test that measures average
glucose levels over time), which includes his time using \system, was
5.8\%. That is nearly non-diabetic levels of control -- healthy
individuals will have an A1C of 5.6\% or lower. \emph{All} of the
students decreased the amount of time spent in hypoglycemia.  These
results are profound for the seven people using \system, and it had a
positive impact on their health and quality of life.

Finally, Bob was a willing and enthusiastic user of the \system
system. To be clear, Bob is \emph{not} a graduate student who was
forced to use \system so that his advisor could publish a
paper. Rather, Bob struggled with the cognitive load of T1D management
and wanted an automated insulin delivery system to offload part of the
burden. However, he was unwilling to use any of the other automated
insulin delivery systems out there due to concerns over their
security. Bob found that \system, with its focus on security,
correctness, and simplicity, met his needs.

\section{Derving the implementational invariant}
\label{sec:invariant_derivation}
In this section, we show how we derive the implementational invariant
from the biological invariant. We introduce the following equations to
capture the error bounds on CGMs and pumps under standard conditions:
\begin{equation}
  0.9 \cdot G_{t}^{A} \leq G_{t}^{M} \leq 1.1 \cdot G_{t}^{A} \tag{3}
\end{equation}
\begin{equation}
0.99 \cdot I_{t}^{A} \leq I_{t}^{M} \leq 1.01 \cdot I_{t}^{A} \tag{4}
\end{equation}
Here, $I_{t}^{M}$ and $G_{t}^{M}$ denote the measured insulin dose and
glucose concentration respectively at time $t$, while $I_{t}^{A}$ and
$G_{t}^{A}$ represent their true values.

We now recall the biological invariant (Equation $(1)$ in Section \ref{sec:bioinvariant}):
\begin{align}
\Delta G^{C} - \Delta G^{A} &\leq T, \quad \text{or} \quad \Delta G^{C} - T \leq \Delta G^{A} \tag{5}
\end{align}

$\Delta G^{A} = G_{t+\Delta t}^{A} - G_{t}^{A}$
is the actual drop in glucose over the same period. We note that we
do not know the value of $\Delta G^{A}$ since we cannot capture the
user's true glucose concentration.

Here, $\Delta G^{C}$ is the calculated drop in the user's glucose over
period $\Delta t$ and $T$ is the clinical threshold of 30 mg/dl.

We adapt mathematical models from prior literature \cite{simglucose,
  physio2019} to compute $\Delta G^{C}$:

\begin{equation}
  \Delta G^{C} = S \cdot (B \cdot \Delta t - \Delta IOB^{A} + \Sigma I_{i}^{A}) \tag{6}
\end{equation}

where $S$ is the \kernel's estimate of the user's insulin sensitivity and $B$ is
the user's basal rate (a physiological parameter representing the amount of
insulin required per hour to absorb the background glucose produced by the body).
Basal rate does not change as frequently as the insulin sensitivity and often
remains stable for days \cite{Lindmeyer2021BasalProfiles}. $\Delta IOB^{A}$ captures
the change in the amount of insulin in the user's body over the period $\Delta t$ and
$\Sigma I_{i}^{A}$ is the sum of insulin doses delivered over the same period.
$\Delta IOB^{A}$ at time $t$ is calculated as the dot product of insulin doses
across the 6-hour window ending at $t$ with coefficients from standard
exponential curves. In other words, both $\Delta IOB^{A}$ and
$\Sigma I_{i}^{A}$ depend on insulin doses which cannot be captured precisely.

To account for these unknowns, we combine $(3)$, $(4)$, $(5)$ and $(6)$ to get:

\begin{equation}
S \leq \frac{0.9 \cdot G_{t+\Delta t}^{M} - 1.1 \cdot G_{t}^{M} + T} {B \cdot \Delta t - 0.99 \cdot \Delta IOB^{M} + 0.01 \cdot \Sigma I_{i}^{\
    M}} \tag{7}
\end{equation}

Here, $\Delta IOB^{M}$ and $\Sigma I_{i}^{M}$ are the measured equivalents
of $\Delta IOB^{A}$ and $\Sigma I_{i}^{M}$, recorded from the pump. We note
that Equation $(7)$ is the implementational invariant(Equation $(2)$
in Section \ref{sec:bioinvariant}), thereby showing how we derive
it from the biological invariant.

\begin{figure}[t]
  \centering
  \includegraphics[width=\linewidth]{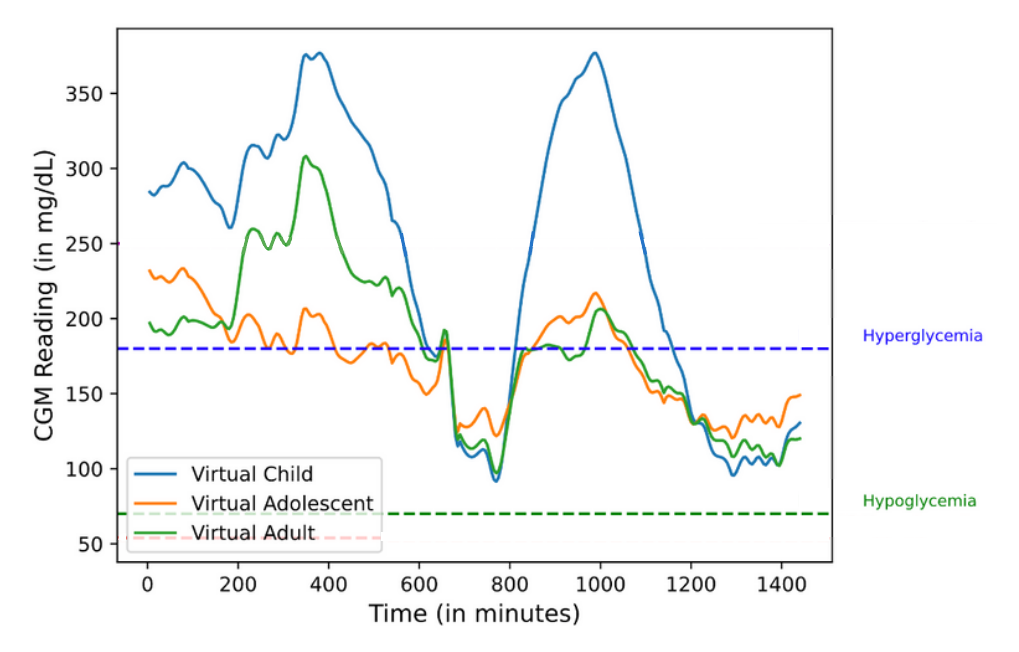}
  \caption{Variation in CGM readings for three virtual humans when
    running a predictive ML algorithm that intentionally doses one-tenth
    the amount of the required insulin. The CGM readings staying mostly
    above the hyperglycemic bound would have had drastic
    health consequences in all three individuals.}
  \label{fig:malicious_less}
  \hrulefill
\end{figure}

\begin{figure}[t]
  \centering
  \includegraphics[width=\linewidth]{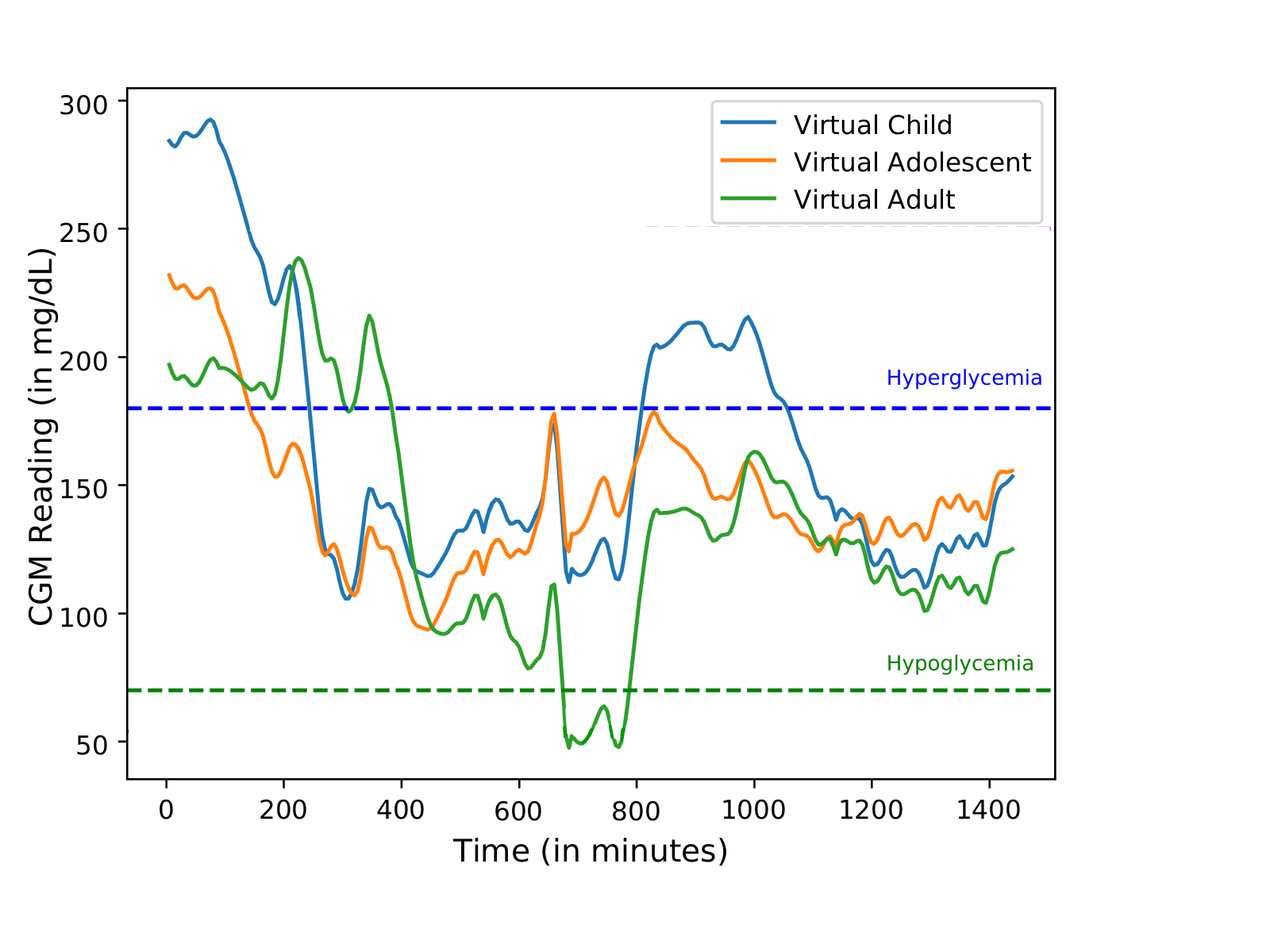}
  \caption{Variation in CGM readings for three virtual humans when clamping an algorithm that
    intentionally tries to dose one-tenth the amount of the required insulin. On average, CGM
    readings stay in range over 70\% of the time.}
  \label{fig:safety_less}
  \hrulefill
\end{figure}

\section{Evaluating algorithmic security when intentionally underdosing insulin}
\label{sec:underdose_eval}
We repeat the experiment described in Section \ref{sec:alg_eval}, but intentionally
dose one-tenth the required amount of insulin. This experiment represents an
algorithm that intentionally underdoses insulin as opposed to overdosing insulin.
Figure \ref{fig:malicious_less} shows that the same virtual humans described in
Section \ref{sec:alg_eval} spend most of their time above the hyperglycemic bound
which would have led to severe health consequences in all three individuals.

Figure \ref{fig:safety_less} shows the same virtual humans in the same simulation
scenarios with the same malicious algorithm dosing one-tenth the required
amount of insulin, but with \system's algorithmic security mechanism enforced.
We now see all virtual humans spending over 70\% of their time in the recommended
healthy range.

\section{Driver security: Withstanding malicious pump drivers}
\label{sec:driver_security}
Pump drivers provide abstractions for insulin delivery applications to
communicate with insulin pumps. Much like traditional device drivers,
they have complex implementations since they help accomplish a wide
range of tasks including pump setup/teardown, managing pump UX,
handling insulin delivery/suspension, managing pump errors, accurately
tracking insulin doses, and so on. 

Given this complexity, there is a strong possibility for bugs or
vulnerabilities to be present in existing pump drivers. These
vulnerabilities can harm or even lead to the death of individuals since
pump drivers control insulin pumps. For example, the user could be under
risk of insulin overdose if the pump driver inadvertently or intentionally
issues the same insulin dosing command twice.

Our goal is to protect users from inadvertent bugs in pump drivers as
well as malicious pump drivers. We introduce a shim between the pump
driver and the pump, which validates all communication between them.
We design our mechanism to interpose on existing pump driver-pump
communication instead of directly programming the pump. This designs
ensures that our trusted computing base remains small and simple by
not including complex pump driver implementations. This design
also provides flexibility with altering or introducing new functionality
to pump drivers.

Interestingly, to protect the individual, we had to overrule our principle
of not programming the pump in cases pertaining to commands for canceling
ongoing insulin delivery. We could not directly adopt the design of
other device driver security mechanisms \cite{ref_validation, tang2010trust,
  expressos} for our driver security mechanism since these cases are specific
to insulin pumps. 

We focus on pump drivers because they are complex (14.5k LoC), multi-threaded,
event-driven, and use two-way communication between the pump and the driver.
In contrast, CGM drivers are simple (1.9k LoC) and only read sensor values
from the device. In our future work, we will formally verify CGM drivers
using the same techniques we use for the insulin delivery path
(Section \ref{sec:formal_verification}).

\subsection{Driver security mechanism design}

\begin{figure}[t]
\centering
\includegraphics[width=\columnwidth]{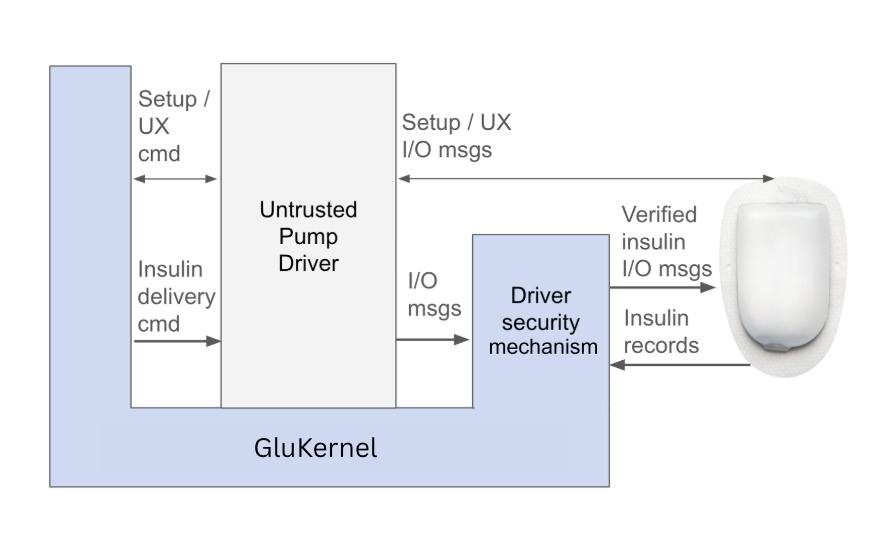}
\caption{We interpose on communications between the pump driver and the pump to 
handle untrusted pump drivers.}
\label{fig:pump_overview}
\hrulefill
\end{figure}

Figure \ref{fig:pump_overview} provides an overview of our driver
security mechanism. The \kernel registers high-level commands, like
``deliver 2U of insulin'' with our driver security mechanism, and sends
them to the pump driver. The pump driver converts these commands into
I/O messages to program the pump. The driver security mechanism then checks
to make sure that these I/O messages are consistent with the previously
registered high-level pump commands before forwarding them to the pump
hardware. We only \validate commands for insulin delivery since they
pose the biggest threats to the user. On detecting added, modified or
dropped commands from the pump driver, the driver security mechanism rejects
further commands and transitions the system to manual insulin delivery where
the user takes control. We list insulin message types pertaining to insulin delivery,
pump setup and UX in Figure \ref{fig:extra_insulin_commands}.

Dropped commands pertaining to canceling ongoing insulin delivery
require separate consideration since excess insulin is dangerous and
cannot be withdrawn once it has been delivered. Users will cancel
insulin delivery if they accidentally issued an insulin delivery
command or incorrectly set the insulin dosage. In cases where a
malicious driver drops a cancel command, the driver security mechanism
directly issues I/O messages to the pump to cancel insulin delivery
before transitioning the pump to a manual mode. We make the tradeoff
of increasing the size of our trusted computing base to handle these
cases in the interest of protecting the user. Fortunately, there are
only two commands to cancel insulin delivery in our current
implementation, so the increase to our trusted computing base is
small.

\begin{figure}[t]
  \centering
  \includegraphics[width=0.75\linewidth]{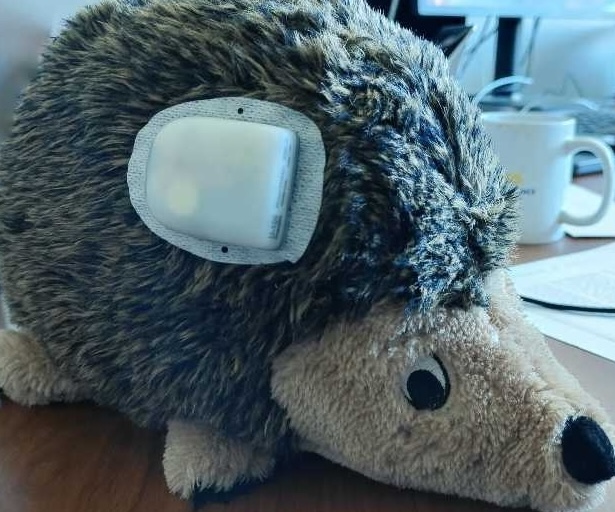}
  \caption{We attached an Omnipod DASH insulin pump to a stuffed animal and modified the OmniB\
LE pump driver to intentionally add, drop and modify insulin commands. Our driver security mec\
hanism detected all manipulations.}
  \label{fig:malicious_driver}
  \hrulefill
\end{figure}

\subsection{Does \system ensure pump operation while overcoming malicious pump drivers?}
\label{sec:driver_eval}
We evaluate the driver security mechanism in allowing communications from
a benign pump driver to go through while blocking manipulated communications
from a malicious pump driver.

\textbf{Malicious pump drivers} We modified the OmniBLE pump driver \cite{omnible}
to intentionally add, drop and modify insulin commands sent to an Omnipod DASH
insulin pump \cite{omnipod} attached to a stuffed animal (Figure
\ref{fig:malicious_driver}). We report that the driver security mechanism
detected all manipulated commands, resulting in 0 false negatives.

\textbf{Benign pump drivers} For one week, we logged messages passed
between the \kernel, the unmodified OmniBLE pump driver and the pump
on an individual running \system (Section \ref{sec:case_study}). We
report 100\% consistency between all commands showing that \system's
driver security mechanism does not incur any false positives.

These results show that \system can ensure pump operation while
withstanding attacks from malicious drivers.

\begin{figure}[t]
    \centering
    \small
    \begin{tabular}{|c|c|c|}
      \hline
      {\bf Message} & {\bf{Message}} & {\bf Message} \\
      {\bf name}  & {\bf{type}} & {\bf description} \\\hline
      {} & {Insulin} & {Request insulin delivery} \\
      {getStatus} & {delivery} & {status from the pump} \\ \hline
      {} & {Insulin} & {Insulin dose delivery} \\
      {statusResponse} & {delivery} & {information from the pump} \\ \hline
      {} & {Insulin} & {Insulin dose delivery} \\
      {errorResponse} & {delivery} & {error from the pump}\\ \hline
      {} & {Insulin} & {Set default rate of} \\
      {setBasalSchedule} & {delivery} & {insulin delivery} \\ \hline
      {} & {Insulin} & {Set temporary rate} \\
      {setTempBasal} & {delivery} & {of insulin delivery} \\ \hline
      {} & {Insulin} & {} \\
      {bolus} & {delivery} & {Deliver instant insulin dose} \\ \hline
      {} & {Insulin} & {} \\
      {cancelDelivery} & {delivery} & {Cancel insulin delivery} \\ \hline
      {setupPod} & {Setup} & {Setup a new pump} \\ \hline
      {assignAddress} & {Setup} & {Pair pump with phone} \\ \hline
      {deactivatePod} & {Setup} & {Teardown pump} \\ \hline
      {acknowledgeAlert} & {UX} & {Acknowledge alert from pump} \\ \hline
      {configureAlerts} & {UX} & {Configure alerts from pump} \\ \hline
      {beepConfig} & {UX} & {Configure pump beeps} \\ \hline
    \end{tabular}
    \caption{Types of messages exchanged between a pump driver and pump pertaining to delivery, setup and UX.}
    \label{fig:extra_insulin_commands}
    \hrulefill
\end{figure}

\begin{figure}[t]
\centering
\includegraphics[width=\columnwidth]{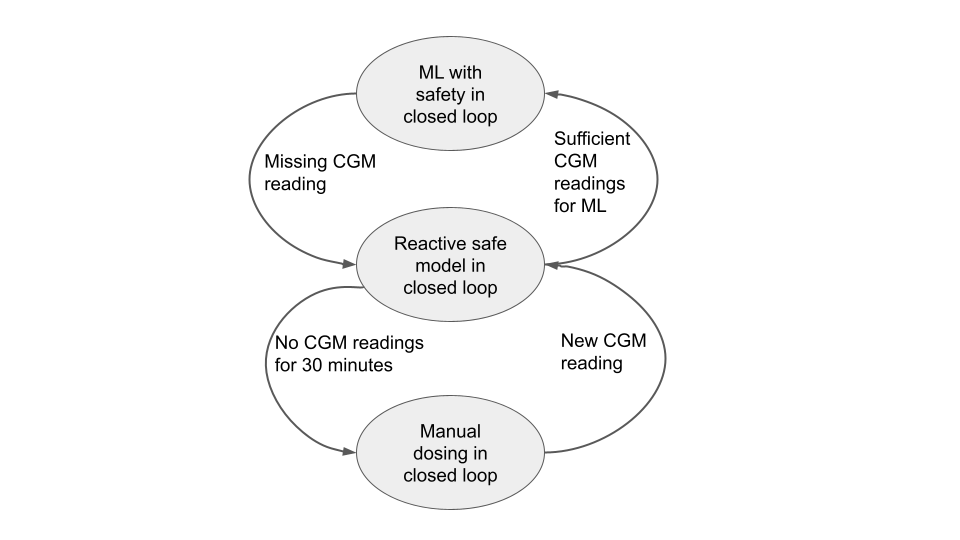}
\caption{\system's CGM state machine to handle missing readings from CGMs.}
\label{fig:cgm_state_machine}
\hrulefill
\end{figure}

\section{Runtime failure management}
\label{sec:runtime_security}

\begin{figure}
\centering
\includegraphics[width=\columnwidth]{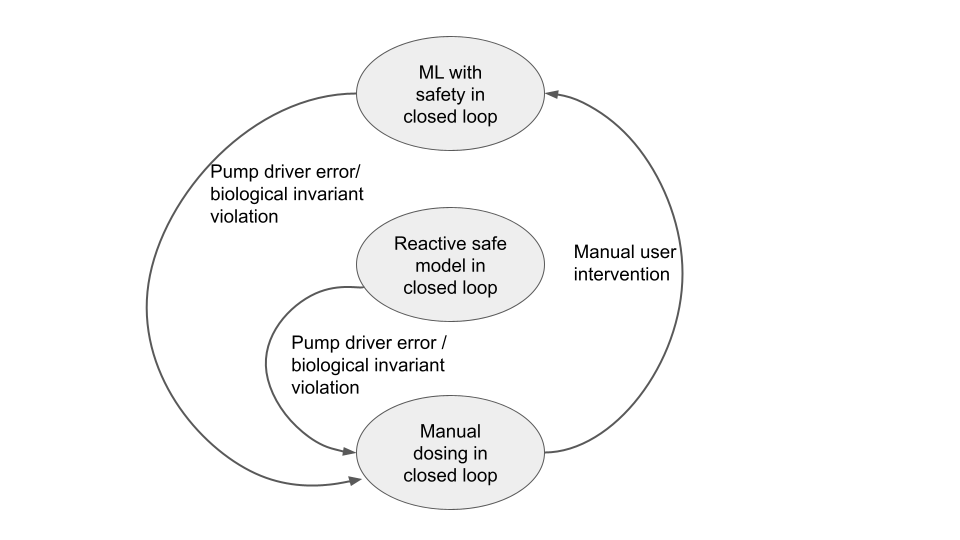}
\caption{\system's state machine to transition users to the safest manual 
dosing state when losing trust in insulin delivery (i.e., with a buggy or 
malicious pump driver) or when the biological invariant is not satisfied.}
\label{fig:state_machine}
\hrulefill
\end{figure}

In this section, we describe \system's \emph{state machine} that
combines its security components, i.e., algorithmic security (Section
\ref{sec:security_design}), driver security (Section
\ref{sec:driver_security}) and end-to-end security (Section
\ref{sec:bioinvariant}) together to protect users in light of their
corresponding failures.

\system's state machine consists of three operating states: (1) ML
with safety operating in closed loop, (2) reactive safe model
operating in closed loop and (3) manual dosing in closed loop. ML
provides the tightest level of control in managing glucose but poses
the largest immediate threat from excess insulin since it preemptively
injects insulin by anticipating rise in glucose. The reactive safe
model provides slightly lesser control, but poses milder immediate
threat since it only injects insulin to account for measured excess
glucose.  Manual dosing offers the least amount of control but poses
no threat since it only injects the minimum insulin needed for
sustenance. \system predominantly operates in the state running ML
(with safety) in closed loop.

We transition to the safest manual dosing state when we do not trust
insulin delivery (i.e., with a buggy or malicious pump driver). We
also transition to the safest manual operation when the biological
invariant is not satisfied to ensure that we do not risk the
possibility of removing too much glucose from our insulin doses.  If
we end up in the manual dosing state, we enforce that only manual user
intervention can transition back to other states. We visualize these
state transitions in Figure \ref{fig:state_machine}.

Inability to retrieve readings from a CGM also count as runtime
failures. However, these failures are easier to
handle.

Since ML needs a stream of CGM readings as input, we transition from
ML to the reactive safe model if we don't receive an input from the
CGM. As long as the number of missing CGM readings is low (say, within
30 minutes), we can interpolate to account for the missing readings to
operate the reactive safe model. Once we have too many missing CGM
readings that make us lose confidence in interpolation, we transition
to manual dosing. From manual dosing, we transition back to the
reactive safe model on receiving a CGM reading. Lastly, from the
reactive safe model, we transition back to ML once we have enough CGM
readings that are required by the ML
model. Figure~\ref{fig:cgm_state_machine} summarizes these
transitions.

\begin{figure*}[t]
    \centering
    \begin{tabular}{|p{5cm}|p{10cm}|}
    \hline
    \multicolumn{1}{|c|}{\textbf{Bug Type}} & \multicolumn{1}{c|}{\textbf{Description}} \\
    \hline
    Division by Zero Error & Potential division by zero error when calculating insulin dosage if the user entered the same start and end time for a basal rate, resulting in a duration of zero. \\
    \hline
    Negative Value Handling & The insulin dosing calculation for microbolusing and basal rate erroneously allowed negative values. \\
    \hline
    Out-of-Range Value Handling & The microbolusing function failed to handle cases where one of its calculation components was out of the expected range or resulted in a negative value. \\
    \hline
    Violation of Intended Postcondition & The function responsible for setting the insulin delivery duration violated its intended postcondition by allowing the end date to be earlier than the start date. \\
    \hline
    Violation of Intended Postcondition & The function responsible for calculating the total insulin dose violated its intended postcondition by returning an incorrect upper bound value under certain conditions. \\
    \hline
    Violation of Intended Postcondition & The function responsible for updating the user's insulin dosage violated its intended postcondition by failing to validate the input parameters correctly. \\
    \hline
    Violation of Intended Postcondition & The function responsible for calculation of ‘unitsDelivered‘ needed additional postconditions accurately reflect the amount of insulin delivered over the time gap. \\
    \hline
    Lack of Error Handling & Additional error handling was added to insulin delivered over a time segment function to gracefully catch undesirable state. \\
    \hline
    Lack of Error Handling & Additional error handling was added to microbolusing function to gracefully catch undesirable states. \\
    \hline
    \end{tabular}
    \caption{Report of bugs fixed by formally verifying \system.}
    \label{fig:bugs_report}
    \hrulefill
\end{figure*}

\section{Formal verification: invariants and bugs}
\label{sec:invariants_list}
In this section, we list the invariants used to formally verify critical
functions of \system. Figure \ref{fig:bugs_report} lists the bugs we found and fixed during formal verification.
\subsection{Invariants}
The MicroBolusing function ensures that micro boluses are always administered at safe time intervals while making sure that they are actually needed (the glucose level of the person is in fact high).

\begin{itemize}
  \item Ensure no micro bolus in the last 4.2 minutes.
  \item Ensure glucose is at least 20 mg/dl above the target glucose level.
  \item Ensure predicted glucose is greater than the current glucose level minus 2.
  \item Ensure micro bolus amount is within the allowed range.
  \item Ensure bolus amount is rounded correctly if \emph{pumpManager} is available.
\end{itemize}

We assume that the \kernel is the only software issuing bolus commands.

The GuardRails function is responsible for ensuring a safe basal rate value, mindful of the user's glucose levels. It enforces several invariants: 
\begin{itemize}
  \item Ensure the result of rounding \emph{newBasalRateRaw} to the supported basal rate is within the allowable range.
  \item Ensure the new basal rate does not exceed the maximum basal rate in the settings.
  \item Ensure the new basal rate is not negative.
  \item Ensure that if either the current glucose level or the predicted glucose level falls below or equals the shut-off glucose threshold, the new basal rate is set to 0.0.
\end{itemize}
The function operates under the assumption that all paths must go through the guard rails, with a specific evaluation comparing with the pump invariant. Additionally, it assumes the integrity of settings, trusting the underlying OS and file system and storage stack.

The insulinDeliveredForSegment function ensures that \emph{intersectionStart} is not greater than \emph{intersectionEnd} when computing the intersection between \emph{self.startDate} and \emph{self.endDate} with \emph{startDate} and \emph{endDate}.
\begin{itemize}
  \item Verifies that \emph{intersectionDuration} is non-negative and does not exceed \emph{doseDuration}, ensuring it accurately reflects the overlapping time duration.
  \item Checks that the units of insulin (\emph{units}) used for calculation are either \emph{deliveredUnits} or \emph{programmedUnits}, ensuring consistency in how the insulin amount is derived.
  \item Verifies that the computation \emph{(units * intersectionDuration / doseDuration)} accurately represents the amount of insulin delivered during the intersection period.
\end{itemize}
The function assumes that the input parameters (\emph{self.startDate}, \emph{self.endDate}, \emph{startDate}, \emph{endDate}, \emph{units}, and \emph{doseDuration}) are valid and within the expected ranges. It is also assumed that the underlying data structures and calculations are accurate and consistent throughout the system.

The CreateBasalDose function requires a time gap greater than 1 second to proceed and return a valid \emph{DoseEntry}.
\begin{itemize}
  \item Ensures that the calculation of \emph{basalRatePerSecond} from \emph{basalRate} accurately represents the rate of insulin delivery per second.
  \item Ensures that the calculation of \emph{unitsDelivered} accurately reflects the amount of insulin delivered over the time gap.
  \item The \emph{DoseEntry} constructed should have consistent attributes and adhere to the specified type, units (\emph{unitsPerHour}), and mutability (\emph{false}).
\end{itemize}
The function assumes that the input parameters (\emph{basalRate}, \emph{startDate}, and \emph{endDate}) are valid and within the expected ranges. It also assumes that the underlying time calculations and conversions between units are accurate and consistent.

The inferBasalDoses function ensures that \emph{basalDoses} contains only \emph{DoseEntry} instances where \emph{type} is \emph{.tempBasal}, \emph{.resume}, or \emph{.suspend}, sorted in ascending order by \emph{startDate}.
\begin{itemize}
  \item InsulinType should be determined correctly based on the last \emph{.tempBasal} or \emph{.bolus} type dose in \emph{doses}, defaulting to \emph{.humalog} if none are found.
  \item Each inferred basal dose added to \emph{inferredBasalDoses} must be created using the \emph{createBasalDose} function with valid parameters and added in the correct chronological order.
  \item If the last dose in \emph{basalDoses} is not of type \emph{.suspend}, a valid basal dose must be inferred from its \emph{endDate} to \emph{at}.
\end{itemize}
The function assumes that the input parameters (\emph{doses}, \emph{basalDoses}, \emph{at}, and \emph{pumpRecordsBasalProfileStartEvents}) are valid and within the expected ranges. It also assumes the correctness of the underlying data structures and the \emph{createBasalDose} function.

The insulinOnBoard function ensures that \emph{doses} contains unique and valid \emph{DoseEntry} instances filtered up to \emph{at}.
\begin{itemize}
  \item The total \emph{iob} should accurately represent the cumulative insulin on board from all valid dose entries in \emph{doses}.
  \item \emph{basalDoses} should contain inferred basal doses that accurately reflect insulin on board contributions when \emph{pumpRecordsBasalProfileStartEvents} is false.
  \item The final return value of \emph{insulinOnBoard} should be the sum of \emph{iob} and \emph{basalIob}, representing the total insulin on board at \emph{at}.
\end{itemize}
The function assumes that the input parameters (\emph{doses}, \emph{at}, and \emph{pumpRecordsBasalProfileStartEvents}) are valid and within the expected ranges. It also assumes the correctness of the underlying data structures and the \emph{inferBasalDoses} function.

The ActiveBolus function ensures that data is read consistently from disk before proceeding with any operations.
\begin{itemize}
  \item Verifies that \emph{DeduplicatedDoses(events: eventLog, at: at)} correctly filters out duplicates and retains only relevant dose entries up to \emph{at}.
  \item Ensures that doses.filter accurately selects bolus entries that were active at the specified \emph{at}.
  \item Verifies that \emph{doses.last} correctly returns the last active bolus entry, ensuring it is non-null if a bolus exists within the specified criteria.
\end{itemize}
The function assumes that the input parameters (\emph{eventLog} and \emph{at}) are valid and within the expected ranges. It also assumes the correctness of the underlying data structures and the \emph{DeduplicatedDoses} function, as well as the consistency and reliability of disk read operations.

The TempBasal (Safety Clamps) function ensures that the events array includes only events from safetyStates that fall within the time range \emph{[start - duration, at)}.
\begin{itemize}
  \item Verifies that the historicalMlInsulin value represents the total units of insulin delivered by the machine learning system over the specified time horizon.
  \item Ensures that the machine learning and safety temp basal rates are correctly converted to units of insulin based on the given duration.
  \item Ensures that the upper and lower bounds for insulin units are correctly adjusted based on historicalMlInsulin.
  \item Ensures that the difference between machine learning and safety temp basal units is correctly clamped within the calculated bounds.
  \item The clamped delta units should be converted back to a temp basal rate and added to the safety temp basal rate.
  \item The returned \emph{SafetyTempBasal} object should correctly encapsulate the adjusted temp basal rate and historical machine learning insulin.
\end{itemize}
The function assumes that the input parameters (\emph{safetyStates}, \emph{start}, \emph{duration}, \emph{at}, \emph{mlTempBasal}, \emph{safetyTempBasal}, and \emph{lastHistoricalMlInsulin}) are valid and within the expected ranges. It also assumes the correctness of the underlying data structures and the conversion functions between units, as well as the proper configuration and appropriateness of the safety clamp parameters and thresholds for the user.

\section{Scenarios when the user's insulin sensitivity is lower than the \kernel's estimate}
\label{sec:low_sensitivity}
We do not take any action when the actual change in the user's glucose
is higher than the calculated change. Higher actual glucose does not
necessarily reflect a shift in the user's physiological state (decreased
insulin sensitivity) since it can also result from digestion when
the user consumes food. Further, even if the user is less sensitive to
insulin, their glucose absorption would be slower, which does not pose
them with an immediate threat (hyperglycemia).

\end{document}